\documentclass[aps,prb,reprint,amsmath,amssymb,longbibliography]{revtex4-2}

\usepackage[T1]{fontenc}
\usepackage[utf8]{inputenc}
\usepackage{microtype}
\usepackage{graphicx}
\graphicspath{{figures/}{../figures/}}
\usepackage{bm}
\usepackage{mathtools}
\usepackage{xcolor}
\usepackage[colorlinks=true,allcolors=blue!55!black]{hyperref}

\newcommand{\ii}{\mathrm{i}}

\newcommand{\dd}{\mathrm{d}}
\newcommand{\uu}{\mathrm{u}}
\newcommand{\AB}{\mathrm{AB}}

\newcommand{\Arg}{\operatorname{arg}}

\newcommand{\cI}{\mathcal{I}}

\begin{document}

\title{Probing quantum Hall edge chirality and the anyonic exchange angle with a three-path interferometer}

\author{Eugene V. Sukhorukov}
\affiliation{Department of Physics, University of Geneva, CH-1211 Geneva, Switzerland}

\date{\today}

\begin{abstract}
An upstream neutral mode can influence quasiparticle tunneling even when
electric charge propagates only downstream.  We propose a three-path
fractional quantum Hall interferometer that probes this directional
structure through the flux dependence of average terminal currents.
Interference between direct and two-step tunneling first contributes at
cubic order in the tunneling amplitudes of three weak quantum point
contacts.  In the local theory, equal drain voltages separate the downstream
and upstream responses: for a purely downstream tunneling excitation, the
cubic fundamental Aharonov--Bohm harmonic vanishes at one drain, while the
other remains bright.
Averaging measurements with exchanged unequal drain voltages extends this
separation to a configuration in which all contacts are biased.  It requires
independent calibration of the electrostatic phase shift and fixed or known
tunneling magnitudes.  For a selected Abelian excitation with two-point
correlation exponent $0<\Delta_{\bm\ell}<1$, the common low-temperature bias
power and directional amplitude ratio determine its scaling dimension and
principal exchange angle.  We also show that a weak static density
interaction spanning two tunneling points can activate the dark harmonic
without adding an upstream mode.  In the high-bias static limit, the
first-order Laughlin activated amplitude scales as $E^{2\nu-1}$, where $E$
is the source--drain bias energy and $\nu$ is the filling factor.  In the weak localized
single-mode infrared regime, the bright harmonic retains its leading
scaling and can serve as a reference in the same device.  The calculation
assumes one coherent tunneling species and common-temperature edge
correlations with unresolved propagation times.
\end{abstract}

\maketitle

\section{Introduction}
\label{sec:introduction}

The fractional quantum Hall effect provides a setting in which charge,
statistics, coherence, and directed transport are encoded in the same
low-energy edge theory~\cite{Laughlin1983,Arovas1984,Wen1990,Wen1992Edges,
FeldmanHalperin2021,Carrega2021}.  Fractional charge was established through
weak-backscattering shot noise~\cite{Saminadayar1997,dePicciotto1997}, while
interferometry and collider experiments probe quasiparticle braiding,
collision phenomena, and scaling
properties~\cite{Nakamura2020,Nakamura2022,Nakamura2023,Bartolomei2020,
Lee2023,Glidic2023,Ruelle2023,VeillonEtAl2024,Werkmeister2025}.  Related
proposals seek access to the exchange angle through single-particle,
collider, antidot, and Hanbury Brown--Twiss
measurements~\cite{SchillerShapiraSternOreg2023,RonettiEtAl2025,
ThammPusterRosenow2026Antidot,PusterThammRosenow2026}.  These approaches
motivate transport measurements that distinguish complementary properties
of a tunneling excitation, including the directional content of its edge
correlations.

The direction of electrical transport does not by itself specify this
content.  We focus on charge-equilibrated edges, where electric charge
propagates downstream and any upstream modes are
neutral~\cite{KaneFisherPolchinski1994,KaneFisher1995Edges,BidEtAl2010,
KumarEtAl2022}.  A quasiparticle operator, or tunneling vertex, can excite
both charged and neutral modes.  Its upstream neutral component can then
influence tunneling at another contact, even though the injected charge is
collected downstream.  Different vertices on the same edge may couple to
different combinations of modes.  The relevant object is therefore the
\emph{selected tunneling excitation}, not the complete edge spectrum.
An electrical probe of this structure must be sensitive to the correlations
between tunneling events at spatially separated contacts.

We use three-path interference to bring these correlations into the
Aharonov--Bohm (AB) oscillations of average terminal currents, complementing
established routes to quantum Hall quasiparticle
spectroscopy~\cite{Chamon1997,LawFeldmanGefen2006,PonomarenkoAverin2007,
LevkivskyiFrohlichSukhorukov2012}.  The device consists of three reservoir-fed
edge segments coupled pairwise by weak quantum point contacts (QPCs), as
shown in Fig.~\ref{fig:setup}.  A direct transfer through one QPC interferes
with a coherent two-step alternative through the other two.  The enclosed
flux enters the product of all three tunneling amplitudes, so the
fundamental AB harmonic first appears at cubic order.  The two time
orderings of the tunneling events probe the response between the contacts
of the intermediate segment in opposite directions.  Their different
dependences on the terminal voltages allow the downstream and upstream
contributions to be separated in the measured current.

The simplest separation uses one source and two drains at equal voltages.
In the local theory, charge balance cancels the downstream contribution at
one drain and the upstream contribution at the other.  A purely downstream
selected excitation therefore gives no cubic fundamental AB harmonic at one
drain, while its neighbor remains bright.  The source is still biased and
all three QPC amplitudes are nonzero.  This dark harmonic is absent at every
flux, rather than only at a tuned interference minimum.  The separation
holds at common finite temperature without a high-bias approximation,
providing a directional test in a single voltage setting.

The separated harmonics can also be used for quantitative spectroscopy
through their bias dependence.  Equal drains, however, leave the contact
between them unbiased, so temperature must keep that contact in the
weak-tunneling regime.  We therefore extend the measurement
to two settings with unequal drain voltages, exchanging those voltages while
holding the source and mean drain voltage fixed.  Before averaging the two
flux traces, we compensate the independently calibrated bias-induced
electrostatic phase shift.  The QPC tunneling magnitudes remain fixed or
independently known, both between the paired settings and along the bias
sweep.  The average then isolates one directional harmonic at each drain,
now with all three contacts biased.

The bias dependence and relative magnitude of these harmonics provide
complementary information about the selected excitation.  On an edge with
counterpropagating modes, the scaling dimension measures its total
correlation weight, whereas the exchange angle depends on the difference
between its directional
weights~\cite{Haldane1995,LevkivskyiBoyarskyFrohlichSukhorukov2009}.  For a
vertex labeled by $\bm\ell$, write these weights as $\delta_+$ and
$\delta_-$, with $\Delta_{\bm\ell}=\delta_++\delta_-=2h_{\bm\ell}$ the
two-point correlation exponent and $h_{\bm\ell}$ the scaling dimension.
At fixed nonzero drain-detuning ratio and temperature small compared with
every QPC bias energy, each nonzero averaged amplitude scales as
$E^{3\Delta_{\bm\ell}-2}$.  Here $E=q_{\bm\ell}(V_S-V_D)$ is the
source-to-mean-drain bias energy; the charge $q_{\bm\ell}$ is independently
calibrated rather than inferred from the flux period of the complete
electronic device~\cite{ByersYang1961,LevkivskyiFrohlichSukhorukov2012}.
The relative amplitudes resolve how the total weight is distributed between
the two directions:
\begin{equation}
 r_{\bm\ell}
 =\frac{A_{\bm\ell}^{\uu}}{A_{\bm\ell}^{\dd}}
 =\left|\frac{\sin(\pi\delta_-)}{\sin(\pi\delta_+)}\right|.
 \label{eq:intro_ratio}
\end{equation}
Here $A_{\bm\ell}^{\uu}$ and $A_{\bm\ell}^{\dd}$ are the upstream and
downstream oscillation amplitudes.  For $0<\Delta_{\bm\ell}<1$, the measured
exponent fixes the sum of the weights, and the ratio determines their
difference, hence the principal exchange angle
$\theta_{\bm\ell}=\pi(\delta_+-\delta_-)$.  The extraction thus combines
scaling and relative amplitudes, rather than identifying a fringe phase
directly with the exchange angle.

The additional directional information is explicit at the
disorder-dominated $\nu=2/3$ fixed point.  A selected charge-$e/3$ excitation
and the neutral-free charge-$2e/3$ excitation have the same scaling dimension,
but amplitude ratios $2$ and $0$, respectively.  The latter gives a dark
upstream harmonic even though the edge still supports an upstream neutral
mode.  The contrast records the participation of that neutral mode in the
selected excitation: in this local example, it is the neutral component
that carries the upstream influence between the contacts.

An upstream influence can also arise from a spatially nonlocal coupling
between the contacts, without an upstream propagating mode.  A density
interaction spanning the QPC pair can join two downstream propagation
processes and activate the dark harmonic even for a purely downstream
excitation.  Returning to equal drain voltages, we evaluate this mechanism
for one weak static interaction on a Laughlin segment.  The leading response
separates into a spatial interaction moment and the net tunneling spectrum
of the other two edges.  For a nonzero moment in the high-bias static regime,
the first-order activated amplitude obeys
\begin{equation}
 A_{a-1}^{\mathrm{dark}}\propto E^{2\nu-1}.
 \label{eq:intro_bridge_results}
\end{equation}
Here $a$ labels the interacting segment.  For $0<\nu<1/2$, the local bright
harmonic instead scales as $E^{\nu-1}T^{2\nu-1}$.  A weak, localized
finite-range coupling in the regular single-mode infrared regime preserves
these leading bright powers at the interaction order considered, while
renormalizing the prefactor.  The bright signal can therefore serve as a
scaling reference in the same device, without switching off the natural
interaction.  These distinct powers characterize the specified nonlocal
mechanism, rather than a universal separation of intrinsic and
interaction-induced contributions.

Both the local spectroscopy and the interaction comparison are developed
within a weak-tunneling description.  The same coherent quasiparticle species
is assumed to dominate all three QPCs, and the stationary edge sectors have
a common temperature.  Bias and temperature lie within the validity range
of the selected edge theory and do not resolve the propagation times of the
participating modes.  The explicit interaction response is retained to first
order in the coupling.

The paper is organized as follows. Section~\ref{sec:setup} defines the device, edge model, and measured current.
Section~\ref{sec:cubic_current} derives the directional response, and
Sec.~\ref{sec:kmatrix} develops the equal-drain test and calibrated
spectroscopy.  Section~\ref{sec:bridge} analyzes the static nonlocal
interaction, followed by discussion and conclusions in
Sec.~\ref{sec:discussion}.  Appendices~\ref{app:cubic_current}--\ref{app:bridge}
supply the supporting derivations.

\section{Three-path device and general Abelian edge model}
\label{sec:setup}

The three-path interferometer is built from three reservoir-fed quantum Hall
edge sectors connected pairwise by weak quantum point contacts (QPCs).  The
part of each sector between its two QPCs remains coherent, while the exterior
continuations are absorbed and re-emitted incoherently by a macroscopic
terminal.  We focus on the charge-equilibrated regime, in which electric
charge propagates downstream and any upstream modes are neutral.  The
tunneling quasiparticle may nevertheless excite both charged and neutral modes,
so its vertex operator need not be chiral.  Each uncoupled sector is assumed to
be in a stationary grand-canonical state at voltage $V_a$ and common temperature
$T$, including its incoming neutral modes.  Weak QPC tunneling leaves these
incident distributions unchanged to the perturbative order considered below.
This section introduces the edge theory, the quasiparticle selected at the
QPCs, the oriented tunneling Hamiltonian, and the conserving current expansion.

\subsection{Geometry and terminal observables}
\label{subsec:geometry_observables}

\begin{figure}[t]
 \centering
 \includegraphics[width=0.82\columnwidth]{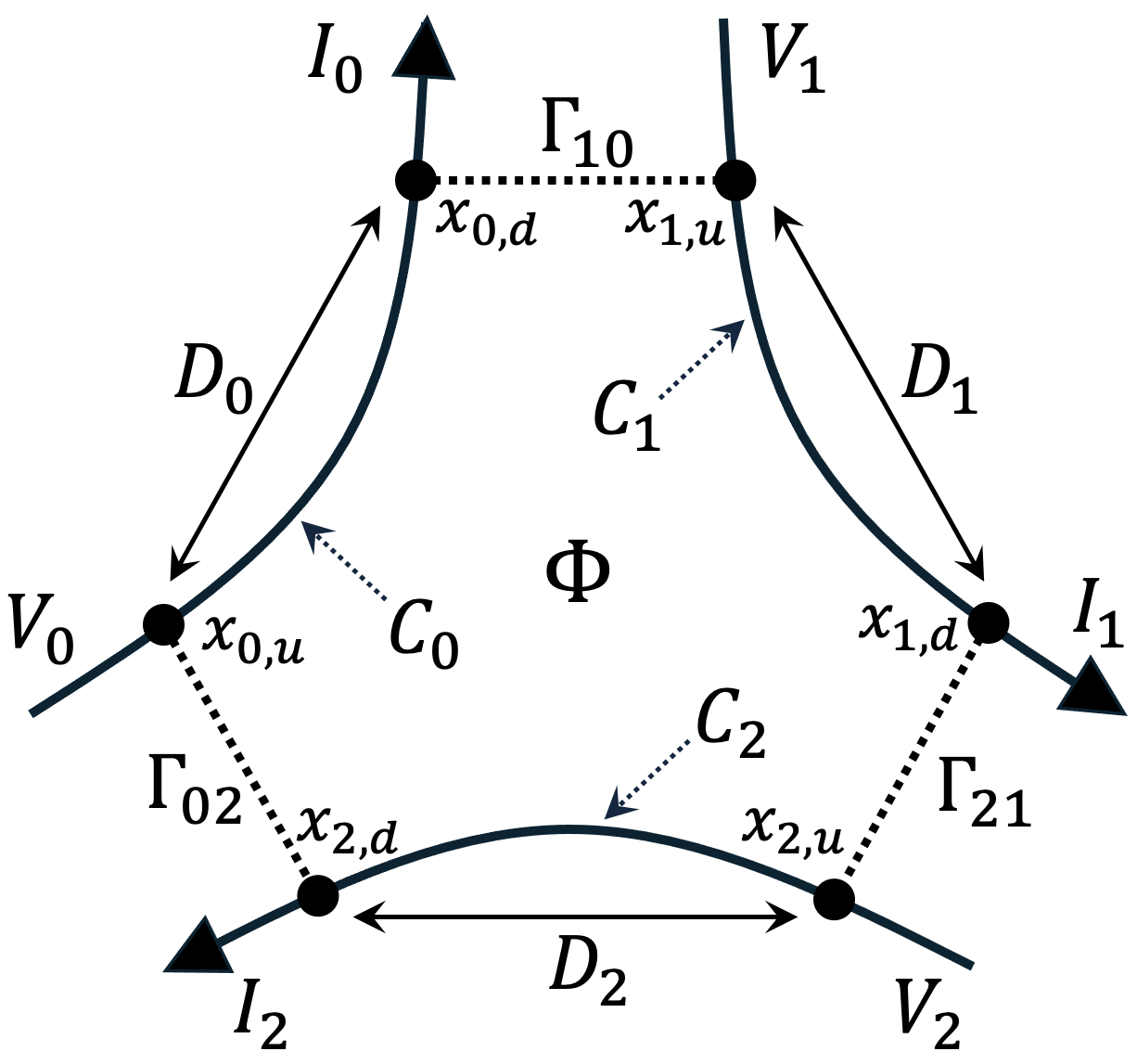}
 \caption{Three-path interferometer.  The arrows fix the positive
 coordinate and downstream electrical propagation on each coherent segment;
 upstream modes, when present, are neutral.  The modes incident on $C_a$ are
 prepared by terminal $a$ at voltage $V_a$ and common temperature $T$.
 The measured current $I_a$ is the excess downstream current into that terminal,
 relative to the uncoupled device at the same voltages and temperature.}
 \label{fig:setup}
\end{figure}

Figure~\ref{fig:setup} shows the coherent edge segments $C_a$, with
$a\in\mathbb Z_3$ and all indices understood modulo three.  We choose one
laboratory coordinate on each segment that increases in the nominal downstream
charged direction,
\begin{equation}
 x_{a,u}\longrightarrow x_{a,d},
 \qquad
 D_a=x_{a,d}-x_{a,u}>0.
 \label{eq:II_coordinate_orientation}
\end{equation}
A propagation eigenmode with velocity $v_j>0$ travels from $x_{a,u}$ to
$x_{a,d}$, whereas a mode with $v_j<0$ travels in the opposite direction.  The
QPC joining $x_{a-1,d}$ to $x_{a,u}$ is assigned the directed amplitude
\begin{equation}
 \Gamma_{a,a-1}:\quad C_{a-1}\longrightarrow C_a,
 \label{eq:II_qpc_orientation}
\end{equation}
and the Hermitian-conjugate term describes the reverse transfer.  Thus
$\Gamma_{10}$, $\Gamma_{21}$, and $\Gamma_{02}$ form the positive oriented
cycle $0\to1\to2\to0$.

We define $I_b$ as the excess downstream electrical current flowing into
reservoir $b$, relative to the uncoupled device at the same voltages and
temperature.  The incident electrical current is unchanged, so in the
stationary state this excess equals the net tunneling current into sector $b$,
\begin{equation}
 I_b=\left\langle\left.\dot Q_b\right|_T\right\rangle.
 \label{eq:II_terminal_current_definition}
\end{equation}
Here $Q_b$ is the total charge of the reservoir-fed sector, and
$\left.\dot Q_b\right|_T$ counts charge transferred into it through its two
QPCs.  Thus $I_b>0$ means that extra positive charge is collected by reservoir
$b$.

The tunneling Hamiltonian only redistributes charge among the three sectors, so
$\sum_b I_b=0$ at every perturbative order.  Coherence is required between the
two QPCs on a given $C_a$, not through a reservoir.  The open geometry
therefore has no dynamical winding-number sector, although the closed product
of the three QPC amplitudes remains sensitive to the enclosed magnetic flux.

\subsection{General Abelian edge and selected quasiparticle}
\label{subsec:abelian_model}

The device is intended to characterize the quasiparticle species selected by
the QPCs.  We therefore formulate the local edge theory in the general Abelian
$K$-matrix language from the outset.  The three coherent segments are assumed
to realize the same low-energy fixed point.  In terms of the field vector
$\boldsymbol\phi_a$, the uncoupled Hamiltonian, equal-time algebra, and positive
charge density are
\begin{align}
 H_0&=\frac{\hbar}{4\pi}\sum_a\int dx\,
 \partial_x\boldsymbol\phi_a^{T}\mathsf V_{\rm edge}
 \partial_x\boldsymbol\phi_a,
 \label{eq:II_edge_hamiltonian}\\
 [\partial_x\phi_{aI}(x),\phi_{bJ}(y)]
 &=2\pi\mathrm{i}\,\delta_{ab}(K^{-1})_{IJ}\delta(x-y),
 \label{eq:II_edge_algebra}\\
 \rho_a(x)&=\frac{e}{2\pi}\boldsymbol t^{T}
 \partial_x\boldsymbol\phi_a(x).
 \label{eq:II_edge_density}
\end{align}
Here $K$ is symmetric, integral, and nondegenerate, while
$\mathsf V_{\rm edge}$ is real, symmetric, and positive definite.  It fixes
velocities and interactions and is unrelated to the terminal voltages $V_a$.
The vector $\boldsymbol t$ specifies the electromagnetic charge.  These are the
standard data of an Abelian quantum Hall edge
\cite{Wen1990,Wen1992Edges,Haldane1995,FrohlichZee1991}.

Away from the QPCs, the corresponding signed current density is
\begin{equation}
 j_a(x,t)=-\frac{e}{2\pi}\boldsymbol t^T
 \partial_t\boldsymbol\phi_a(x,t).
 \label{eq:II_edge_continuity}
\end{equation}
Taking $j_a$ positive along increasing $x$, the measured excess at a
cross-section $x_{a,d}^{\rm cut}$ downstream of both QPCs is
\begin{equation}
 I_a=\bigl\langle j_a(x_{a,d}^{\rm cut})\bigr\rangle
       -\bigl\langle j_a(x_{a,d}^{\rm cut})\bigr\rangle_0.
 \label{eq:II_terminal_cross_sections}
\end{equation}
The subscript $0$ denotes the uncoupled reference at the same reservoir
voltages and temperature.  Upstream neutral modes may affect tunneling at
another QPC, but they carry none of the injected electric charge upstream.
Edges with unequilibrated counterpropagating charged modes lie outside this
charge-collection description.

We choose one integer vector $\boldsymbol\ell$ and assume that the corresponding
quasiparticle tunnels at all three QPCs.  Its annihilation operator on $C_a$ is
\begin{align}
 \mathcal V_{\boldsymbol\ell,a}(x)
 &=:
   \!\exp[\mathrm{i}\boldsymbol\ell^T\boldsymbol\phi_a(x)]\!:
 \nonumber\\[-1mm]
 &=:
   \!\exp\!\left[\mathrm{i}\sum_jp_j\varphi_{aj}(x)\right]\!:
 .
 \label{eq:II_selected_vertex}
\end{align}
The two expressions describe the same operator in the original $K$-matrix
basis and in the propagation-eigenmode basis.  Our sign convention is that
$\mathcal V_{\boldsymbol\ell,a}$ annihilates a positive quasiparticle and
$\mathcal V_{\boldsymbol\ell,a}^{\dagger}$ creates one.  Its charge and Abelian
exchange angle are
\begin{equation}
 q_{\boldsymbol\ell}
 =e\,\boldsymbol t^TK^{-1}\boldsymbol\ell>0,
 \qquad
 \theta_{\boldsymbol\ell}
 =\pi\boldsymbol\ell^TK^{-1}\boldsymbol\ell
 \pmod{2\pi}.
 \label{eq:II_charge_exchange}
\end{equation}
The charge is treated as independently calibrated, for example by a weak-QPC
shot-noise measurement.  We do not infer it from an unspecified thermodynamic
flux period of the complete electronic device.

To expose propagation direction, choose a real matrix $M$ such that
\begin{align}
 \boldsymbol\phi_a&=M\boldsymbol\varphi_a,
 \qquad
 M^TKM=\Sigma=\operatorname{diag}(\sigma_j),
 \nonumber\\[-1mm]
 M^T\mathsf V_{\rm edge}M&=\operatorname{diag}(u_j),
 \label{eq:II_mode_normalization}
\end{align}
where $\sigma_j=\pm1$, $u_j>0$, and $v_j=\sigma_j u_j$.  The coefficients in
Eq.~\eqref{eq:II_selected_vertex} are
$\boldsymbol p=M^T\boldsymbol\ell$.  Our electrical-transport assumption
requires $(M^T\boldsymbol t)_j=0$ for $v_j<0$, but does not require $p_j=0$
for those neutral modes.  We define the downstream and upstream weights of
the selected vertex by
\begin{align}
 \delta_+&=\sum_{v_j>0}p_j^2,
 &\delta_-&=\sum_{v_j<0}p_j^2,
 \nonumber\\[-1mm]
 \Delta_{\boldsymbol\ell}&=\delta_++\delta_-,
 &h_{\boldsymbol\ell}&=\frac{\Delta_{\boldsymbol\ell}}{2}.
 \label{eq:II_directional_weights}
\end{align}
Here $\Delta_{\boldsymbol\ell}$ is the full two-point-function exponent and
$h_{\boldsymbol\ell}$ is the conventional scaling dimension.  The weights
$\delta_\pm$ describe vertex correlations, not fractions of the electric
charge.  Since $K^{-1}=M\Sigma M^T$,
\begin{equation}
 \theta_{\boldsymbol\ell}
 =\pi(\delta_+-\delta_-)
 \pmod{2\pi}.
 \label{eq:II_sum_difference}
\end{equation}
Thus the correlator exponent fixes the sum of the directional weights, whereas
the exchange angle fixes their difference.  A purely downstream vertex has
$\delta_-=0$ even if other operators of the same edge contain upstream modes.
For the elementary Laughlin quasiparticle at $\nu=1/m$,
\begin{equation}
 q_{\boldsymbol\ell}=\nu e,
 \qquad
 \delta_+=\Delta_{\boldsymbol\ell}=\nu,
 \qquad
 \delta_-=0,
 \qquad
 \theta_{\boldsymbol\ell}=\pi\nu.
 \label{eq:II_Laughlin_specialization}
\end{equation}
The normalized mode construction is given in
Appendix~\ref{app:B_modes_ordering}.

\subsection{Oriented tunneling Hamiltonian and loop phase}
\label{subsec:tunneling_loop}

The directed operator on link $a-1\to a$ is
\begin{equation}
 \mathcal A_{a,a-1}
 =\mathcal V_{\boldsymbol\ell,a}^{\dagger}(x_{a,u})
  \mathcal V_{\boldsymbol\ell,a-1}(x_{a-1,d}),
 \qquad C_{a-1}\longrightarrow C_a.
 \label{eq:II_directed_tunneling_operator}
\end{equation}
It transfers one positive quasiparticle from terminal sector $a-1$ to $a$.
Short-distance normalization is absorbed into the energy-valued QPC amplitudes
$\Gamma_{a,a-1}$.  A physical transfer $b\to c$ releases the energy
\begin{equation}
 \epsilon_{bc}=q_{\boldsymbol\ell}(V_b-V_c),
 \qquad
 \epsilon_{01}+\epsilon_{12}+\epsilon_{20}=0.
 \label{eq:II_oriented_energy_drop}
\end{equation}
In the voltage-phase representation, the uncoupled operators evolve with
$K_0=H_0-\sum_bV_bQ_b$, and expectation values $\langle\cdots\rangle_0$ are
taken with $\rho_0=Z_0^{-1}\exp(-K_0/T)$.  The tunneling Hamiltonian is
\begin{align}
 H_T(t)&=\sum_a\left[
 \mathcal T_{a,a-1}(t)+\mathcal T_{a,a-1}^{\dagger}(t)
 \right],
 \nonumber\\
 \mathcal T_{a,a-1}(t)&=
 \Gamma_{a,a-1}
 \exp[-\mathrm{i}\epsilon_{a-1,a}t/\hbar]\,
 \mathcal A_{a,a-1}(t).
 \label{eq:II_tunneling_hamiltonian}
\end{align}
The explicit exponential accounts for the electrochemical energy released
by each tunneling event.  Appendix~\ref{app:cubic_current} derives the cubic
current directly in this voltage-phase representation.

The positive oriented loop is
\begin{equation}
 \Gamma_{\circlearrowright}
 =\Gamma_{02}\Gamma_{21}\Gamma_{10}
 =|\Gamma_{\circlearrowright}|\exp(\mathrm{i}\varphi_\Gamma).
 \label{eq:II_loop_product}
\end{equation}
Operators on distinct coherent segments are assigned one fixed global order,
as specified in Appendix~\ref{app:B_modes_ordering}.  No independent dynamical
Klein-factor sector is introduced for the neutral cubic average-current loop.
In this convention the flux-dependent loop phase is
\begin{equation}
 \varphi_{\mathrm{AB},\boldsymbol\ell}
 =2\pi\frac{q_{\boldsymbol\ell}}{e}\frac{\Phi}{\Phi_0}
 +\varphi_\Gamma,
 \qquad \Phi_0=\frac{h}{e}.
 \label{eq:II_loop_phase}
\end{equation}
We normalize the cubic fundamental harmonic as
\begin{equation}
 I_b^{(3),\mathrm{AB}}(\Phi)
 =\frac{2q_{\boldsymbol\ell}|\Gamma_{\circlearrowright}|}{\hbar^3}
 \operatorname{Re}\!\left[
 e^{\mathrm{i}\varphi_{\mathrm{AB},\boldsymbol\ell}}
 \mathcal I_b
 \right].
 \label{eq:II_harmonic_definition}
\end{equation}
The reduced complex coefficient $\mathcal I_b$ contains the voltage,
temperature, and propagation dependence.  The phases generated by the edge
response remain in $\mathcal I_b$; they are distinct from a change of the
common QPC or electrostatic phase between voltage settings.  The statistical
factors associated with the two time orderings of tunneling events at the
contacts bounding an intermediate segment are derived in
Sec.~\ref{sec:cubic_current} and Appendix~\ref{app:vertex_loop}.  Their relative
exchange phase remains explicit in the response.  Reversing the loop
orientation complex conjugates the complete positive-loop coefficient.

To connect the coefficient explicitly to a flux sweep, write the fundamental
fringe in amplitude--phase form,
\begin{equation}
 I_b^{(3),\mathrm{AB}}(\Phi)
 =A_b\cos\!\left[\varphi_{\mathrm{AB},\boldsymbol\ell}+\vartheta_b\right],
 \label{eq:II_harmonic_polar}
\end{equation}
where
\begin{equation}
 A_b=\frac{2q_{\boldsymbol\ell}|\Gamma_{\circlearrowright}|}{\hbar^3}
          |\mathcal I_b|,
 \qquad \vartheta_b=\arg\mathcal I_b.
 \label{eq:II_amplitude_phase}
\end{equation}
The nonnegative current amplitude $A_b$ is half the peak-to-peak variation
of this harmonic.  The response phase $\vartheta_b$ describes its displacement
relative to the common loop phase in Eq.~\eqref{eq:II_loop_phase}; it is not,
in general, the exchange angle $\theta_{\boldsymbol\ell}$.  No response phase
is assigned when $A_b=0$.  These definitions also apply to an averaged
fundamental fringe once its complex coefficient has been formed.  Section
\ref{sec:cubic_current} now evaluates $\mathcal I_b$ as the sum of physical
downstream and upstream terminal contributions.

\subsection{Perturbative regime and conserving method}
\label{subsec:regime_method}

The calculation assumes one dominant quasiparticle species and a common local
edge fixed point and temperature on the three segments.  The low-energy edge
theory applies when all bias energies and the temperature obey
\begin{equation}
 E_{\rm exp}
 =\max\!\left(T,|\epsilon_{01}|,|\epsilon_{12}|,|\epsilon_{20}|\right)
 \ll E_{\rm uv}.
 \label{eq:II_low_energy_window}
\end{equation}
Here $E_{\rm uv}$ is the upper validity scale of the selected edge theory
and its vertex normalization.  When using an infrared fixed point, as in the
disorder-dominated $\nu=2/3$ example of Sec.~\ref{subsec:K_23_example}, this
scale must lie below the crossover into that fixed-point regime, not merely
below the bulk gap.  Weak tunneling is a separate, contact-dependent
requirement: the renormalized amplitude on link
$a-1\to a$ must be small at its own infrared scale
$\max\{T,|\epsilon_{a-1,a}|\}$.  For a relevant tunneling operator,
this scale must remain above that contact's strong-coupling crossover energy,
$T_{B,a}$.  In particular, temperature must keep an unbiased contact weak;
a large voltage drop on another contact does not supply its infrared cutoff.

Neither temperature nor bias is allowed to resolve the flight time between the
two QPCs on any coherent segment:
\begin{equation}
 \max_{a,j,b,c}\left\{
 \frac{2\pi T D_a}{\hbar|v_j|},
 \frac{|\epsilon_{bc}|D_a}{\hbar|v_j|}
 \right\}\ll1.
 \label{eq:II_unresolved_flight}
\end{equation}
The maximum includes every mode participating in the tunneling vertex,
including neutral modes.  The commutators are evaluated for spatially
separated QPCs before the short propagation delays are neglected.  This
preserves which contact can influence the other, even when the flight time is
unresolved.  A short separation $D_a$ does not by itself establish the
fixed-point regime required in Eq.~\eqref{eq:II_low_energy_window}.

The net tunneling current in Eq.~\eqref{eq:II_terminal_current_definition}
is generated by the charge commutator.  Since the uncoupled Hamiltonian
conserves $Q_b$, the transfer convention in
Eq.~\eqref{eq:II_directed_tunneling_operator} gives
\begin{align}
 \widehat I_b(t)
 &=\left.\dot Q_b(t)\right|_T
   =\frac{\mathrm{i}}{\hbar}[H_T(t),Q_b]
 \nonumber\\[-1mm]
 &=-\frac{\mathrm{i}q_{\boldsymbol\ell}}{\hbar}
 \sum_a(\delta_{ba}-\delta_{b,a-1})
 \left[\mathcal T_{a,a-1}(t)
      -\mathcal T_{a,a-1}^{\dagger}(t)\right].
 \label{eq:II_current_operator}
\end{align}
A transfer $a-1\to a$ gives a positive contribution to $I_a$ and a negative
one to $I_{a-1}$.  The opposite signs at the two ends of each link give the
operator identity $\sum_b\widehat I_b=0$.  Since the current operator is
already first order in tunneling, the cubic current follows by expanding the normalized real-time
contour to second order in $H_T$:
\begin{align}
 \langle I_b(0)\rangle^{(3)}
 ={}&-\frac{1}{\hbar^2}
 \int_{-\infty}^{0}dt_1
 \int_{-\infty}^{t_1}dt_2
 \nonumber\\[-1mm]
 &\times\left\langle
 [[\widehat I_b(0),H_T(t_1)],H_T(t_2)]
 \right\rangle_0.
 \label{eq:II_nested_commutator}
\end{align}
An adiabatic switch-on is understood.  Equation~\eqref{eq:II_nested_commutator} is the standard real-time nested-commutator expansion
\cite{Kamenev2011,Rammer2007}.  It contains all current-insertion positions and contour orderings required by
the normalized closed-time evolution; its derivation is given in Appendix~\ref{app:cubic_current}.
The shortest gauge-invariant flux-sensitive term uses each QPC once, so the
fundamental AB harmonic first appears in this cubic contribution.  Section III
now resolves this complete result according to whether the response crosses an
intermediate segment downstream or upstream.

\section{Cubic Aharonov--Bohm current and directional response}
\label{sec:cubic_current}

The fundamental loop harmonic uses each QPC once and therefore first appears
in the cubic nested-commutator expression
\eqref{eq:II_nested_commutator}.  This formulation is particularly useful in
a chiral system because the direction of propagation is already fixed by the
inner commutator.  We first identify its downstream and upstream responses and
then give the resulting terminal current.  Appendix~A provides the complete
real-time derivation, while Appendix~\ref{app:vertex_loop} evaluates the
Abelian vertex algebra and the common loop correlator.

\subsection{Propagation between the two QPCs}
\label{subsec:cubic_direction}

Consider the contribution to Eq.~\eqref{eq:II_nested_commutator} from the
part of the current operator associated with the QPC $C_{a-1}\to C_a$.
This operator is evaluated at $t=0$, while the two remaining QPC operators
of the positive loop act at earlier times, $t_2<t_1<0$.  Their two possible
time orderings give different causal responses.  In the first, the inner
commutator connects the upstream QPC of $C_{a-1}$ at $t_1$ to its downstream
QPC at $0$.  In the second, it connects the downstream QPC of $C_a$ at $t_1$
to its upstream QPC at $0$.  These are the downstream and upstream responses,
respectively, whose explicit expressions are given in
Eqs.~\eqref{eq:A_Md} and \eqref{eq:A_Mu}.

\begin{figure*}[t]
 \centering
   \includegraphics[width=0.72\textwidth]{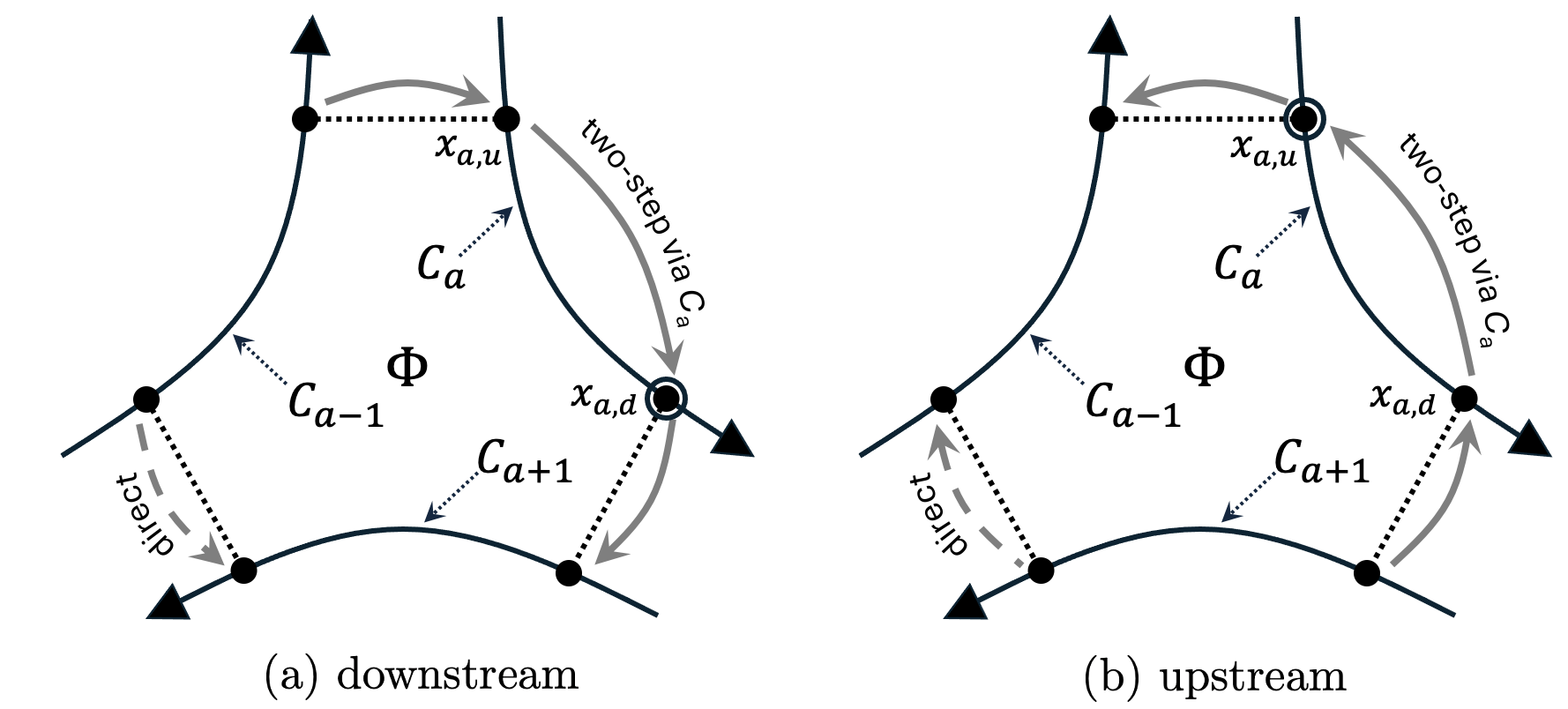}
 \caption{
Spatial illustration of the two causal orderings through a selected
segment $C_a$ in the cubic Aharonov--Bohm current.
The dark edge arrows fix the nominal downstream orientation, and the
black dotted links denote the QPCs.
The dashed gray arrow shows the direct one-QPC alternative, while the
solid gray arrows show the two-QPC alternative through $C_a$.
(a) The causal response connects $x_{a,u}$ to $x_{a,d}$.
(b) The response instead connects $x_{a,d}$ to $x_{a,u}$.
An outer ring marks the QPC associated with the current operator at $t=0$,
the latest time in the nested-commutator expansion: at $x_{a,d}$ in (a)
and $x_{a,u}$ in (b).
For a purely downstream selected vertex in the local theory, the
response in (b) vanishes before time integration.
The panels illustrate causal contributions, not complete terminal
currents; the latter combine the current contributions from the two
contacts adjacent to each terminal.
}
 \label{fig:causal_paths}
\end{figure*}

This statement follows directly from the support of the edge commutators.  It
is useful to write
\begin{align}
 \mathcal A_{a,a-1}(t)&=:\!e^{\mathrm{i}\Theta_a(t)}\!:,
 \nonumber\\[-1mm]
 \Theta_a(t)&=\sum_jp_j
 \left[
 \varphi_{a-1,j}(x_{a-1,d},t)-
 \varphi_{a,j}(x_{a,u},t)
 \right].
 \label{eq:III_Theta}
\end{align}
For $t_1<0$, the two inner commutators have the support
\begin{align}
 [\Theta_a(0),\Theta_{a-1}(t_1)]
 &=2\pi\mathrm{i}\sum_{v_j>0}p_j^2
 \Theta_{\rm H}\!\left(-t_1-\frac{D_{a-1}}{|v_j|}\right),
 \label{eq:III_down_support}\\
 [\Theta_a(0),\Theta_{a+1}(t_1)]
 &=2\pi\mathrm{i}\sum_{v_j<0}p_j^2
 \Theta_{\rm H}\!\left(-t_1-\frac{D_a}{|v_j|}\right).
 \label{eq:III_up_support}
\end{align}
The first line contains only modes with positive velocity and the second only
modes with negative velocity.  In the electrical regime of Sec.~\ref{sec:setup},
the latter response is carried by neutral modes, not by upstream transport of
the injected charge.  The step functions also retain the finite flight-time
threshold.  The short delays are neglected under condition
\eqref{eq:II_unresolved_flight} only after the commutators have been evaluated
at the spatially separated QPCs.

For a purely downstream selected vertex,
\begin{equation}
 \delta_-=\sum_{v_j<0}p_j^2=0,
 \label{eq:III_pure_downstream}
\end{equation}
so every coefficient $p_j$ of a negative-velocity mode vanishes.  The second
commutator in Eq.~\eqref{eq:III_up_support}, and hence the entire upstream
nested contribution, is then zero pointwise before either time integration is
performed.  The third tunneling operator closes the three-QPC loop but cannot
restore a response whose inner commutator vanishes.

For comparison at a fixed segment $C_a$, the current operator is associated
with the QPC at $x_{a,d}$ for the downstream response and with the QPC at
$x_{a,u}$ for the upstream response, as indicated in Fig.~\ref{fig:causal_paths}.
These contributions therefore involve different QPCs.  The measured terminal
combination is formed next.

\subsection{Terminal current}
\label{subsec:terminal_result}

The complex coefficient $\mathcal I_b$ entering the measured harmonic in
Eq.~\eqref{eq:II_harmonic_definition} is the sum of the two propagation
directions described above,
\begin{equation}
 \mathcal I_b=\mathcal I_b^{\mathrm d}+\mathcal I_b^{\mathrm u}.
 \label{eq:III_terminal_decomposition}
\end{equation}
Here $\mathcal I_b^{\mathrm d}$ is generated by a response travelling from the
upstream to the downstream QPC of the intermediate segment, whereas
$\mathcal I_b^{\mathrm u}$ requires propagation in the opposite direction.
Each quantity is a complete directional contribution to the current at
terminal $b$: the contributions from the current operator on its two adjacent
QPCs have already been combined, as in Eq.~\eqref{eq:A_terminal_components}.
It is not the
response through one segment taken in isolation.  Their complex coefficients
add before the magnitude in Eq.~\eqref{eq:II_amplitude_phase} is taken;
the observed fringe therefore depends on both their magnitudes and phases.

Introduce the cyclic QPC drops
\begin{equation}
 \varepsilon_a\equiv\epsilon_{a,a+1},
 \qquad
 \varepsilon_0+\varepsilon_1+\varepsilon_2=0.
 \label{eq:III_cyclic_drops}
\end{equation}
For the ordered events $t_2<t_1<0$, let $u=-t_1>0$ and
$v=t_1-t_2>0$ be the intervals between successive tunneling events.  Each of
the three independent edge sectors contributes a vertex pair correlation,
at a time separation $u$, $v$, or $u+v$.  After the ordering phases have been
extracted, the product of their thermal decay envelopes is the real weight
$W_{\Delta_{\boldsymbol\ell}}(u,v;T)$.  In the common local theory it obeys
\begin{equation}
 W_{\Delta_{\boldsymbol\ell}}(u,v;T)
 =W_{\Delta_{\boldsymbol\ell}}(v,u;T).
 \label{eq:III_kernel_symmetry}
\end{equation}
Its precise form is given by Eqs.~\eqref{eq:B_thermal_envelope} and
\eqref{eq:B_W_definition}.  The voltage-independent complex reference
factor $A_{\Delta_{\boldsymbol\ell}}$ is defined in
Eq.~\eqref{eq:B_factorization}; it is not the measured oscillation amplitude
$A_b$ of Eq.~\eqref{eq:II_amplitude_phase}.  In the unresolved-flight regime,
with current positive into the collecting reservoir as in
Eq.~\eqref{eq:II_terminal_current_definition}, the two terminal contributions are
\begin{align}
 \mathcal I_b^{\mathrm d}
 ={}&\alpha_{\mathrm d}A_{\Delta_{\boldsymbol\ell}}
 \int_0^\infty\!du\int_0^\infty\!dv\,
 W_{\Delta_{\boldsymbol\ell}}(u,v;T)
 \nonumber\\[-1mm]
 &\times\left[
 e^{-\mathrm{i}\varepsilon_bu/\hbar
    +\mathrm{i}\varepsilon_{b+1}v/\hbar}
 -e^{-\mathrm{i}\varepsilon_{b-1}u/\hbar
    +\mathrm{i}\varepsilon_bv/\hbar}
 \right],
 \label{eq:III_terminal_d}\\[1mm]
 \mathcal I_b^{\mathrm u}
 ={}&e^{\mathrm{i}\theta_{\boldsymbol\ell}}\alpha_{\mathrm u}A_{\Delta_{\boldsymbol\ell}}
 \int_0^\infty\!du\int_0^\infty\!dv\,
 W_{\Delta_{\boldsymbol\ell}}(u,v;T)
 \nonumber\\[-1mm]
 &\times\left[
 e^{-\mathrm{i}\varepsilon_bu/\hbar
    +\mathrm{i}\varepsilon_{b-1}v/\hbar}
 -e^{-\mathrm{i}\varepsilon_{b-1}u/\hbar
    +\mathrm{i}\varepsilon_{b+1}v/\hbar}
 \right].
 \label{eq:III_terminal_u}
\end{align}
The two exponentials in each square bracket come from the current-operator
terms on the two QPCs adjacent to terminal $b$.  Their difference enforces charge
balance at that terminal and removes the common energy-independent contact
term at $u=v=0$.  The directional factors $\alpha_{\mathrm d}$ and
$e^{\mathrm{i}\theta_{\boldsymbol\ell}}\alpha_{\mathrm u}$ resolve the
exchange factors of the two causal directions; their explicit expressions
are given below.  The integrals carry the bias and temperature dependence.
They still depend on the total exponent $\Delta_{\boldsymbol\ell}$,
whereas the prefactors also distinguish its downstream and upstream weights.
The relative factor $e^{\mathrm{i}\theta_{\boldsymbol\ell}}$ follows from
ordering the ordinary three-QPC product before evaluating the commutators;
its derivation is in Eqs.~\eqref{eq:B_up_ordered_product}--\eqref{eq:B_factorization}.

We retain the general bias dependence in this integral form because the
symmetry \eqref{eq:III_kernel_symmetry} makes directional separation
possible without first evaluating the integrals.  Section~\ref{sec:kmatrix}
uses it for equal and exchanged drain voltages, then considers the bias
scaling of the surviving harmonics.  The projected scalar integrals are
introduced in Sec.~\ref{subsec:K_directional_separation};
Appendix~\ref{app:B_scalar_support} establishes their nonvanishing and
fixed-detuning scaling limit.

\subsection{Directional coefficients and the local chiral null}
\label{subsec:directional_coefficients}

The ratios of the directional discontinuities to the downstream reference
are
\begin{align}
 \alpha_{\mathrm d}
 &=e^{-\mathrm{i}\pi\delta_-}
   \frac{\sin(\pi\delta_+)}{\sin(\pi\Delta_{\boldsymbol\ell})},
 \label{eq:III_alpha_d}\\
 \alpha_{\mathrm u}
 &=e^{-\mathrm{i}\pi\delta_+}
   \frac{\sin(\pi\delta_-)}{\sin(\pi\Delta_{\boldsymbol\ell})}.
 \label{eq:III_alpha_u}
\end{align}
Appendix~\ref{app:B_exchange_algebra} derives these factors by expanding each
nested commutator into four products of tunneling operators and using the
exchange relations of the quasiparticle vertices.  They are ratios of exchange
coefficients; the relative ordinary-product phase in
Eq.~\eqref{eq:III_terminal_u} is separate.  For a purely downstream vertex, Eq.~\eqref{eq:III_pure_downstream} gives
\begin{equation}
 \alpha_{\mathrm u}=0,
 \qquad
 \mathcal I_b^{\mathrm u}=0.
 \label{eq:III_upstream_null}
\end{equation}
The support argument in Eq.~\eqref{eq:III_up_support} is stronger than the
second equality: it shows that the upstream nested commutator vanishes before
thermal averaging and integration.  A purely negative-velocity vertex
formally gives $\alpha_{\mathrm d}=0$; for a charged vertex this is a
reversed-chirality limit outside the downstream electrical regime assumed here.

\subsection{Conservation and perturbative scope}
\label{subsec:cubic_checks}

Summing Eqs.~\eqref{eq:III_terminal_d} and
\eqref{eq:III_terminal_u} over the three terminals gives
\begin{equation}
 \sum_b\mathcal I_b=0,
 \qquad
 \sum_b I_b^{(3),\mathrm{AB}}=0,
 \label{eq:III_conservation}
\end{equation}
because every QPC current enters once with each sign.  At full equilibrium all
three $\varepsilon_a$ vanish, so every square bracket in
Eqs.~\eqref{eq:III_terminal_d} and \eqref{eq:III_terminal_u} vanishes
separately.  The conjugate loop in
Eq.~\eqref{eq:II_harmonic_definition} makes the measured current real.

The physical terminal differences are universal for
$0<\Delta_{\boldsymbol\ell}<1$, although an individual un-subtracted QPC
contribution is ultraviolet convergent only for
$\Delta_{\boldsymbol\ell}<2/3$.  Appendix~\ref{app:vertex_loop} shows that the
leading contact term cancels in the measured terminal current and that the
distributional limit $\Delta_{\boldsymbol\ell}\to1^-$ reproduces the exact
free-electron Landauer result.

Finally, gauge invariance constrains the fundamental AB harmonic to QPC orders
\begin{equation}
 N=3,5,7,\ldots .
 \label{eq:III_order_scope}
\end{equation}
There is no fourth-order correction.  The cubic null in
Eq.~\eqref{eq:III_upstream_null} is therefore separated from the first
network-mediated background by two additional effective QPC amplitudes; that
background may first appear at fifth order.

\section{Selected-vertex spectroscopy}
\label{sec:kmatrix}

The directional decomposition of the terminal current provides a way to
probe the mode content of the selected tunneling excitation.  We first
consider equal drain voltages, for which a purely downstream vertex
produces a dark fundamental AB harmonic in a single bias setting.  We then
extend this separation to finite drain detuning, where all three QPCs
remain biased, and use the bias dependence and relative magnitude of the
two directional harmonics to determine the vertex's scaling dimension
and principal exchange angle.

\subsection{Equal-drain bias and the dark-terminal harmonic}
\label{subsec:K_equal_drains}

Choose terminal $b-1$ as the source and hold the two drains at the same
voltage:
\begin{equation}
 \begin{gathered}
 V_{b-1}=V_S,\qquad V_b=V_{b+1}=V_D,\\
 E=q_{\bm\ell}(V_S-V_D)>0.
 \end{gathered}
 \label{eq:K_equal_drains}
\end{equation}
For a purely downstream selected vertex, $\delta_-=0$, this configuration
gives
\begin{equation}
 I_b^{(3),\AB}(\Phi)=0
 \qquad\text{for every }\Phi.
 \label{eq:K_equal_drain_null}
\end{equation}
The neighboring drain $b+1$ retains a bright harmonic.  Thus the device
has a dark terminal even though the source remains biased and all three
QPC tunneling amplitudes are nonzero.

Two distinct ingredients produce this null.  First, the equal-drain
voltages cancel the downstream contribution at terminal $b$.  The cyclic
energy drops in Eq.~\eqref{eq:III_cyclic_drops} become
$(\varepsilon_{b-1},\varepsilon_b,\varepsilon_{b+1})=(E,0,-E)$, so the
voltage-dependent difference in Eq.~\eqref{eq:III_terminal_d} reduces to
\begin{equation}
 e^{-\ii Ev/\hbar}-e^{-\ii Eu/\hbar}.
 \label{eq:K_equal_drain_difference}
\end{equation}
This expression is antisymmetric under $u\leftrightarrow v$, whereas the
common thermal weight $W_{\Delta_{\bm\ell}}(u,v;T)$ is symmetric.
The complete integral therefore vanishes.  The upstream contribution at
terminal $b+1$ cancels in the same way, leaving
$\cI_b=\cI_b^{\uu}$ and $\cI_{b+1}=\cI_{b+1}^{\dd}$.
Second, for a purely downstream vertex, local causality makes
$\cI_b^{\uu}=0$, as established in Eq.~\eqref{eq:III_upstream_null}.
Together, these statements give Eq.~\eqref{eq:K_equal_drain_null}.

The equal-drain cancellation is a balance in the net terminal current,
rather than a destructive-interference minimum of a transmission
amplitude.  The two terms in Eq.~\eqref{eq:III_terminal_d} arise from the
current-operator terms at the two QPCs adjacent to the measured reservoir-fed
sector.  They enter with opposite signs because the collected excess current
counts the net charge injected into that sector.  At equal drain voltages, the local equilibrium
correlations give these contributions equal dynamical weights; the
symmetry under $u\leftrightarrow v$ expresses this balance.  Each term
already contains the cubic interference contribution.  Their cancellation
does not require the direct and two-step amplitudes themselves to cancel
at a particular flux.

The free-electron limit, $\nu=1$, makes this mechanism especially
transparent.  The result derived in Appendix~\ref{app:vertex_loop},
Eq.~\eqref{eq:B_Landauer}, is
\begin{equation}
 I_b^{(3),\AB}(\Phi)
 =\frac{2e}{h}\int_{-\infty}^{\infty}d\omega\,
 \operatorname{Re}\Lambda(\omega)
 \bigl[f_b(\omega)-f_{b+1}(\omega)\bigr].
 \label{eq:K_equal_drain_fermion}
\end{equation}
Here $f_c(\omega)=[e^{(\omega-eV_c)/T}+1]^{-1}$ is the occupation of
states incident from reservoir $c$, and $\Lambda$ is the dimensionless
product of the three weak QPC scattering amplitudes and the accumulated
loop phase.  Its flux dependence is implicit.

At cubic order, transmission from the source $b-1$ into terminal $b$
has no AB correction.  The flux-dependent corrections to reflection from
$b$ and transmission from $b+1$ into $b$ are equal and opposite, as
required by unitarity.  Since the two drains have identical incident
occupations, $f_b=f_{b+1}$, these corrections cancel in the terminal
current.  Interference redistributes equally occupied incident states
without producing a net current correction at this terminal.  At the
other drain, the same formula contains the occupation difference between
drain $b+1$ and source $b-1$, which supplies the bright reference.
For fractional quasiparticles, the cancellation follows from the
many-body terminal expressions above, rather than from a single-particle
occupation formula.

The separation holds at common finite temperature without a high-bias
approximation, within the local, same-vertex, weak-tunneling and
unresolved-flight regime of Sec.~\ref{sec:setup}.  For
$0<\Delta_{\bm\ell}<1$ and $E>0$, the bright-harmonic amplitude is
nonzero whenever the QPC product is nonzero; the scalar nonvanishing result
is established in Appendix~\ref{app:B_scalar_support}.  The null characterizes the selected
excitation, not the entire edge theory, and concerns only the cubic
fundamental harmonic.  The dark terminal can still carry a
flux-independent current; higher tunneling orders and spatially
nonlocal interactions can also generate an AB signal there.

\subsection{Directional separation at finite drain detuning}
\label{subsec:K_directional_separation}

Equal drain voltages provide a directional test in a single bias setting,
but leave the QPC joining the drains unbiased.  For relevant quasiparticle
tunneling, temperature must stop the low-energy growth of that coupling
before it reaches strong tunneling; a large source bias does not provide
a voltage cutoff on this link.  Temperature can consequently remain an
essential scale in the leading bright response.  To access a
low-temperature scaling regime with every contact biased, we separate the
drain voltages and recover directional separation by averaging two
settings in which those voltages are exchanged.

Keep the source voltage and the mean drain voltage fixed, and reverse
the drain imbalance:
\begin{equation}
 \mathsf V_\pm:\quad
 \left\{
 \begin{aligned}
 V_{b-1}&=V_S,\\
 V_b&=V_D\pm\frac{\delta V}{2},\\
 V_{b+1}&=V_D\mp\frac{\delta V}{2}.
 \end{aligned}
 \right.
 \label{eq:K_mirror_settings}
\end{equation}
Writing $\varepsilon_D=q_{\bm\ell}\delta V$, we choose
$0<|\varepsilon_D|<2E$.  The source then lies above both drains, and none
of the three QPC voltage drops is zero.  Their magnitudes can be chosen above
the respective strong-coupling crossover scales while the temperature
remains smaller than every bias energy.

The cancellation at equal drains extends to this pair through a reflection
symmetry of the terminal-current expressions.  Exchanging the drain voltages
transforms the cyclic drops as
\begin{equation}
 (\varepsilon_{b-1},\varepsilon_b,\varepsilon_{b+1})
 \longrightarrow
 (-\varepsilon_{b+1},-\varepsilon_b,-\varepsilon_{b-1}).
 \label{eq:K_drop_reflection}
\end{equation}
In the downstream contribution at $b$, this transformation interchanges the
two terms of the terminal difference after $u$ and $v$ are interchanged.
Symmetry of the thermal weight therefore gives a change of sign.  The same
argument applies to the upstream contribution at the neighboring drain:
\begin{equation}
 \begin{aligned}
 \cI_b^{\dd}(\mathsf V_-)&=-\cI_b^{\dd}(\mathsf V_+),\\
 \cI_{b+1}^{\uu}(\mathsf V_-)&=-\cI_{b+1}^{\uu}(\mathsf V_+).
 \end{aligned}
 \label{eq:K_reflection_odd}
\end{equation}
Thus these two unwanted contributions cancel upon averaging the settings,
even though they need not vanish in either setting separately.

Experimentally, the fundamental current oscillation is recorded as a
function of flux at each bias setting.  Before averaging, the two traces are
brought to a common phase reference by compensating the independently
calibrated bias-induced electrostatic phase shift.  The QPC tunneling
magnitudes are kept unchanged or independently normalized.  Averaging the
fundamental oscillations at the same calibrated flux phase is then
equivalent to averaging their complex coefficients:
\begin{equation}
 \cI_c^{\rm sym}
 =\frac{\cI_c(\mathsf V_+)+\cI_c(\mathsf V_-)}{2},
 \qquad c=b,b+1.
 \label{eq:symmetric_harmonic}
\end{equation}
The complex average retains both the magnitude and phase of each fringe.
Averaging positive oscillation amplitudes instead would lose the relative
sign responsible for cancellation.

Using Eq.~\eqref{eq:K_reflection_odd} in the average gives
\begin{equation}
 \begin{aligned}
 \cI_b^{\rm sym}
 &=\tfrac12\bigl[\cI_b^{\uu}(\mathsf V_+)
                  +\cI_b^{\uu}(\mathsf V_-)\bigr],\\
 \cI_{b+1}^{\rm sym}
 &=\tfrac12\bigl[\cI_{b+1}^{\dd}(\mathsf V_+)
                  +\cI_{b+1}^{\dd}(\mathsf V_-)\bigr].
 \end{aligned}
 \label{eq:K_surviving_averages}
\end{equation}
The averaging acts only on the voltage-dependent exponentials in
Eqs.~\eqref{eq:III_terminal_d}--\eqref{eq:III_terminal_u}; the directional
factors, thermal weight $W_{\Delta_{\bm\ell}}$, and normalization
$A_{\Delta_{\bm\ell}}$ are unchanged.  We denote the resulting two-time
integrals, including this normalization, by
$F_{\Delta_{\bm\ell}}^{\dd,\uu}$.  For the downstream contribution at
$b+1$, averaging Eq.~\eqref{eq:III_terminal_d} and interchanging $u$ and
$v$ in one of the two settings gives
\[
\begin{aligned}
 F_{\Delta_{\bm\ell}}^{\dd}(E,\varepsilon_D;T)
 ={}&A_{\Delta_{\bm\ell}}\int_0^\infty\!du\int_0^\infty\!dv\,
 W_{\Delta_{\bm\ell}}(u,v;T)\\
 &\times\Bigl[
 e^{-\ii\varepsilon_{b+1}u/\hbar
       +\ii\varepsilon_{b-1}v/\hbar}\\
 &\qquad-\tfrac12 e^{-\ii\varepsilon_{b-1}u/\hbar
                         +\ii\varepsilon_bv/\hbar}\\
 &\qquad-\tfrac12 e^{-\ii\varepsilon_bu/\hbar
                         +\ii\varepsilon_{b+1}v/\hbar}
 \Bigr].
\end{aligned}
\]
Here the cyclic drops on the right are evaluated in the $\mathsf V_+$
setting.  The upstream function is obtained by the same averaging of the
integral in Eq.~\eqref{eq:III_terminal_u} at drain $b$.  Thus
\begin{equation}
 \begin{aligned}
 \cI_b^{\rm sym}
 &=e^{\ii\theta_{\bm\ell}}\alpha_{\uu}
   F_{\Delta_{\bm\ell}}^{\uu}(E,\varepsilon_D;T),\\
 \cI_{b+1}^{\rm sym}
 &=\alpha_{\dd}F_{\Delta_{\bm\ell}}^{\dd}(E,\varepsilon_D;T).
 \end{aligned}
 \label{eq:K_mirror_recall}
\end{equation}
No further dynamical approximation is involved.  These functions contain
the bias and temperature dependence and depend on the selected vertex
through its total exponent $\Delta_{\bm\ell}$; the separate directional
weights remain in the prefactors.  This separation will allow the common
bias scaling and relative amplitude to be used independently below.

The finite-detuning average therefore recovers the directional separation of
the equal-drain configuration while keeping all three contacts biased.  For
$0<\Delta_{\bm\ell}<1$, the scalar functions are nonzero throughout
the chosen bias domain.  With a nonzero QPC product, a purely downstream
vertex, $\delta_-=0$, consequently gives a dark projected harmonic at $b$
and a bright reference at $b+1$.  This distinction concerns the selected
tunneling excitation, not the complete mode content of its edge.

Both the single-setting null and the finite-detuning separation hold at
finite common temperature without a high-bias approximation.  They require
the common local vertex correlations and stationary reservoir preparation
of Sec.~\ref{sec:setup}, weak tunneling on each link, and the unresolved
flight times in Eq.~\eqref{eq:II_unresolved_flight}.  They concern the
cubic fundamental harmonic, not all tunneling orders.  The different roles
of the two bias configurations will be important for the spectroscopy below
and the nonlocal-interaction example in Sec.~\ref{sec:bridge}.

\subsection{Bias scaling and the scaling dimension}
\label{subsec:K_scaling}

The directional weights defined in Eq.~\eqref{eq:II_directional_weights}
enter the correlator exponent through their sum,
$\Delta_{\bm\ell}=\delta_++\delta_-$, and the exchange angle through
their difference.  The two projected harmonics provide these two pieces of
information separately.  Their common bias dependence determines the sum;
their relative magnitude then determines how that weight is distributed
between the two directions.  This extends edge scaling
spectroscopy~\cite{LevkivskyiBoyarskyFrohlichSukhorukov2009} to the
reconstruction of the principal exchange angle.

Applying Eq.~\eqref{eq:II_amplitude_phase} to the averaged fringes defines
their nonnegative current amplitudes,
\begin{equation}
 \begin{aligned}
 A_{\bm\ell}^{\uu}
 &=\frac{2q_{\bm\ell}|\Gamma_{\circlearrowright}|}{\hbar^3}
   |\cI_b^{\rm sym}|,\\
 A_{\bm\ell}^{\dd}
 &=\frac{2q_{\bm\ell}|\Gamma_{\circlearrowright}|}{\hbar^3}
   |\cI_{b+1}^{\rm sym}|.
 \end{aligned}
 \label{eq:K_projected_amplitudes}
\end{equation}
They are amplitudes of the complex average, not averages of the two fringe
amplitudes.  Increase the voltage differences in a fixed proportion, so
that $\rho=\varepsilon_D/E$ is held fixed away from $0$ and $\pm2$.  In the low-temperature limit each of the three vertex
correlators supplies a factor with exponent $\Delta_{\bm\ell}$, while
the two time integrations contribute two inverse powers of energy.  Thus
each nonzero projected amplitude obeys
\begin{equation}
 A_{\bm\ell}^{\dd},A_{\bm\ell}^{\uu}
 \propto E^{3\Delta_{\bm\ell}-2}.
 \label{eq:K_scaling_exponent}
\end{equation}
This law requires $T\ll E_{\min}$, where
\begin{equation}
 E_{\min}=\min\!\left\{|\varepsilon_D|,
                      E-\frac{|\varepsilon_D|}{2}\right\}
 \label{eq:protocol_min_drop}
\end{equation}
is the smallest QPC bias energy.  The log--log slope determines
$\Delta_{\bm\ell}$ by adding two and dividing by three; the conventional
scaling dimension is then $h_{\bm\ell}=\Delta_{\bm\ell}/2$ as in
Eq.~\eqref{eq:II_directional_weights}.  The common QPC magnitude and vertex
normalization must remain fixed or independently known along the sweep,
not only matched within each voltage pair.

Let $E_{\rm QPC}=\max_a T_{B,a}$ be the largest of the contact crossover
energies defined in Sec.~\ref{subsec:regime_method}.  A sufficient window
for this fixed-temperature scaling measurement is
\begin{equation}
 E_{\rm QPC}\ll E_{\min},\qquad T\ll E_{\min},\qquad
 \max_{b,c}|\epsilon_{bc}|\ll E_{\rm uv}.
 \label{eq:protocol_intrinsic_window}
\end{equation}
The flight-time condition \eqref{eq:II_unresolved_flight} supplies a further
upper bound on the bias.  Thus even a zero or negative power in
Eq.~\eqref{eq:K_scaling_exponent} describes an intermediate-energy
weak-tunneling regime, not the limit $E\to0$ at fixed relevant QPC couplings.

At fixed temperature, Eq.~\eqref{eq:K_scaling_exponent} is a high-bias
asymptotic, not an exact power law through the thermal crossover.  Exact
homogeneity rescales temperature together with all voltage differences;
at fixed nonzero detuning ratio the low-temperature coefficient is finite
and nonzero.  A dark direction does not prevent the exponent measurement:
the bright harmonic still determines $\Delta_{\bm\ell}$.

\subsection{Directional weights and the exchange angle}
\label{subsec:K_inference}

Complex conjugation of the scalar integrals, followed by
$u\leftrightarrow v$, gives
\[
 F_{\Delta_{\bm\ell}}^{\uu}(E,\varepsilon_D;T)
 =e^{\ii\pi\Delta_{\bm\ell}}
  [F_{\Delta_{\bm\ell}}^{\dd}(E,\varepsilon_D;T)]^*.
\]
Their magnitudes are therefore equal at any common temperature.  The common
scalar magnitude and QPC product cancel from the ratio of the two projected
amplitudes, leaving
\begin{equation}
 r_{\bm\ell}
 =\frac{A_{\bm\ell}^{\uu}}{A_{\bm\ell}^{\dd}}
 =\frac{|\cI_b^{\rm sym}|}{|\cI_{b+1}^{\rm sym}|}
 =\left|\frac{\sin(\pi\delta_-)}{\sin(\pi\delta_+)}\right|.
 \label{eq:K_directional_ratio}
\end{equation}
Unlike the fixed-temperature bias power, this relation does not require the
high-bias limit.  It is a ratio of coherent harmonic amplitudes, not a
probability, and can exceed unity.

At fixed $0<\Delta_{\bm\ell}<1$, the ratio in
Eq.~\eqref{eq:K_directional_ratio} increases monotonically with the upstream
weight $\delta_-$.  The measured exponent and amplitude ratio therefore
determine a unique exchange angle on the principal interval
$-\pi\Delta_{\bm\ell}\leq\theta_{\bm\ell}\leq\pi\Delta_{\bm\ell}$:
\begin{equation}
 \tan\frac{\theta_{\bm\ell}}{2}
 =\frac{1-r_{\bm\ell}}{1+r_{\bm\ell}}
  \tan\frac{\pi\Delta_{\bm\ell}}{2}.
 \label{eq:K_theta_reconstruction}
\end{equation}
The directional weights follow from
\begin{equation}
 \delta_\pm=\frac12\left(
 \Delta_{\bm\ell}\pm\frac{\theta_{\bm\ell}}{\pi}\right).
 \label{eq:K_weight_reconstruction}
\end{equation}
The endpoint $r_{\bm\ell}=0$ corresponds to a purely downstream vertex.
The opposite endpoint $r_{\bm\ell}\to\infty$ is the formal purely upstream
limit; a charged vertex of this type lies outside the downstream electrical
regime assumed here.  If one
direction is below experimental sensitivity, the bright exponent and a
bound on the amplitude ratio instead constrain the missing directional
weight.  Near $\Delta_{\bm\ell}=1$ and equal directional amplitudes,
small changes in the ratio can produce large changes in the inferred angle.
The generic inversion is restricted to the stated domain; integer/contact
points require the regulated treatment in Appendix~\ref{app:B_contact_limit}.

The relative response phase of the two averaged fringes, defined as in
Eq.~\eqref{eq:II_amplitude_phase}, has a different role.  When both
signals are nonzero, Eq.~\eqref{eq:K_mirror_recall} gives
\begin{align}
 \Arg\!\left(\frac{\cI_b^{\rm sym}}{\cI_{b+1}^{\rm sym}}\right)
 &=\pi\Delta_{\bm\ell}
 \nonumber\\[-1mm]
 &\quad-2\Arg F_{\Delta_{\bm\ell}}^{\dd}(E,\varepsilon_D;T)
 \pmod{2\pi}.
 \label{eq:K_phase_ledger}
\end{align}
At a given exponent and scaled bias and temperature, this phase tests the
scalar loop response rather than providing another exchange-angle
measurement.  The reconstruction uses the bias exponent and magnitude
ratio.  It determines the aggregate directional weights of the selected
vertex, not the number or velocities of the modes or a unique microscopic
$K$ matrix.

\subsection{Disorder-dominated edge at filling factor
\texorpdfstring{$2/3$}{2/3}}
\label{subsec:K_23_example}

The distinction between scaling and directional information is illustrated
by two excitations of the disorder-dominated $\nu=2/3$ edge.  In the
standard two-component representation,
\begin{equation}
 K=\begin{pmatrix}1&2\\2&1\end{pmatrix},
 \qquad \bm t=\begin{pmatrix}1\\1\end{pmatrix}.
 \label{eq:K_23_data}
\end{equation}
At the Kane--Fisher--Polchinski fixed
point~\cite{KaneFisherPolchinski1994,KaneFisher1995Edges}, the selected
vertex has downstream charge and upstream neutral coefficients
\begin{equation}
 p_c=\frac{\ell_1+\ell_2}{\sqrt6},
 \qquad p_n=\frac{\ell_1-\ell_2}{\sqrt2},
 \label{eq:KFP_modes}
\end{equation}
so that
\begin{equation}
 \frac{q_{\bm\ell}}{e}=\frac{\ell_1+\ell_2}{3},
 \qquad \delta_+=\frac{(\ell_1+\ell_2)^2}{6},
 \qquad \delta_-=\frac{(\ell_1-\ell_2)^2}{2}.
 \label{eq:KFP_weights}
\end{equation}
The vertices in Table~\ref{tab:KFP_vertices} have the same
$\Delta_{\bm\ell}=2/3$ and $h_{\bm\ell}=1/3$, but different
directional content.
\begin{table}[tb]
 \caption{Selected vertices at the disorder-dominated $\nu=2/3$ fixed point.
 Both entries have $\Delta_{\bm\ell}=2/3$ and $h_{\bm\ell}=1/3$.
 The charge-$2e/3$ null refers specifically to the neutral-free
 $\bm\ell=(1,1)$ vertex.}
 \label{tab:KFP_vertices}
 \begin{ruledtabular}
 \begin{tabular}{c c c c c}
 $\bm\ell$ & $q/e$ & $(\delta_+,\delta_-)$ & $\theta/\pi$ & $r$\\
 \hline
 $(1,0)$ or $(0,1)$ & $1/3$ & $(1/6,1/2)$ & $-1/3$ & $2$\\
 $(1,1)$             & $2/3$ & $(2/3,0)$   & $ 2/3$ & $0$
 \end{tabular}
 \end{ruledtabular}
\end{table}
For a selected member of the charge-$e/3$ doublet, the pair
$(\Delta_{\bm\ell},r_{\bm\ell})=(2/3,2)$ gives
$\theta_{\bm\ell}=-\pi/3$ and reveals its upstream neutral weight.
The two doublet members have identical listed invariants and cannot be
distinguished by these measurements.  The neutral-free charge-$2e/3$
vertex instead gives $(2/3,0)$ and $\theta_{\bm\ell}=2\pi/3$.
Its upstream projected harmonic is dark even though the edge still supports
an upstream neutral mode.  Other charge-$2e/3$ operators with neutral
dressing need not obey the same null.

For both entries, $3\Delta_{\bm\ell}-2=0$: the leading nonzero
projected amplitudes are constant along a fixed bias ray in the
low-temperature window.  The bias exponent alone therefore does not
distinguish the two cases, whereas the directional ratio does.  This is an
intermediate-energy weak-tunneling law, not an $E\to0$ asymptote.

The comparison assumes that the indicated coherent species dominates all
three QPCs; equal scaling dimensions do not establish that selection.
The bias and temperature must lie below the crossover into the fixed-point
regime, as required by Eq.~\eqref{eq:II_low_energy_window}, while every
QPC remains weak at its own infrared scale.  The upstream response is
carried by the neutral mode, not by upstream transport of the injected charge.
The stationary reservoir preparation must include the incoming neutral
population, and every participating charge or neutral mode must satisfy
the flight-time condition \eqref{eq:II_unresolved_flight}.  A slow neutral
mode can restrict this window and suppress interference
\cite{GoldsteinGefen2016,BhattacharyyaEtAl2019}.  Under these conditions the
measurement probes the selected excitation within the local edge theory.
Spatial nonlocality provides a different way to connect the two QPCs,
without upstream mode weight, as we now examine.

\section{Static interaction bridge at equal drain voltages}
\label{sec:bridge}

The local response in Sec.~\ref{sec:cubic_current} distinguishes propagation
with and against the edge chirality.  Spatial nonlocality provides a different
way to connect the two QPCs: a density interaction can join two locally
downstream responses without adding an upstream propagating mode.  We study
how this connection produces a fundamental AB harmonic at a terminal whose
bridge-free cubic harmonic is zero.  Throughout this section the same
single-mode downstream Laughlin quasiparticle tunnels at all three QPCs,
with $q=\nu e$ and $0<\nu<1$.  The bright-reference asymptotic below is
restricted to $0<\nu<1/2$, including the elementary fractions
$\nu=1/m$ with odd $m\ge3$.

\subsection{Equal-drain observable and bridge geometry}

For a bridge on $C_a$, return to the equal-drain setting of
Sec.~\ref{subsec:K_equal_drains}, with $b=a-1$ and the selected vertex
specialized to a downstream Laughlin quasiparticle.  As shown below, equal
drains allow us to separate interaction-induced changes in the existing edge
correlations from the new nonlocal response between the QPCs.  Terminal $a+1$
is the source and the other two reservoirs have the same voltage:
\begin{align}
 V_{a+1}&=V_S,\qquad V_a=V_{a-1}=V_D,
 \nonumber\\
 E&=q(V_S-V_D)>0.
 \label{eq:bridge_bias}
\end{align}
The source-first transfer energies of Sec.~\ref{sec:setup} are
\begin{equation}
 \epsilon_{a+1,a-1}=E,\qquad
 \epsilon_{a-1,a}=0,\qquad
 \epsilon_{a,a+1}=-E.
 \label{eq:IV_bridge_drops}
\end{equation}
Thus the QPC $\Gamma_{a,a-1}$ is unbiased.  The equal-drain result of
Sec.~\ref{subsec:K_equal_drains}, specialized to the Laughlin vertex, gives
in the absence of the bridge,
\begin{equation}
 (\mathcal I_{a-1},\mathcal I_a,\mathcal I_{a+1})
 =\bigl(0,F_\nu^{\mathrm d}(E,0;T),-F_\nu^{\mathrm d}(E,0;T)\bigr).
 \label{eq:IV_bridge_reference}
\end{equation}
The zero in the second argument of $F_\nu^{\mathrm d}$ denotes zero drain
detuning.  Terminal $a-1$ is therefore the dark drain and $a$ is the bright
reference.  Here ``dark'' refers to the cubic fundamental harmonic, not to
the full dc current or to every QPC order.

\begin{figure}[t]
 \centering
 \includegraphics[width=0.95\columnwidth]{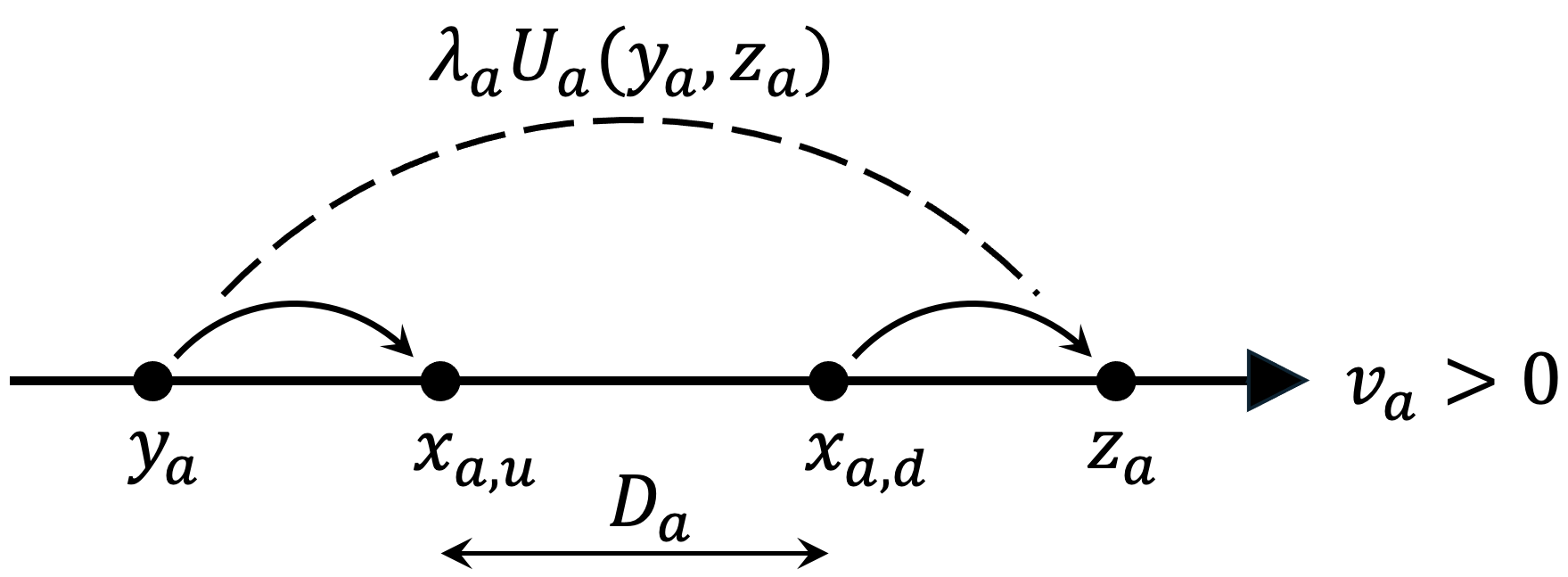}
 \caption{Single interaction bridge on $C_a$.  The solid arrows show
 downstream propagation, and the dashed arc represents the density coupling
 $\lambda_aU_a(y,z)$, not a counterpropagating channel.  By coupling points
 on opposite sides of the QPC pair, the interaction allows the downstream
 QPC to influence the upstream QPC at first order.  The resulting terminal
 harmonic includes contributions from both time orderings of the tunneling
 events.}
 \label{fig:static_bridge}
\end{figure}

With current positive into the collecting reservoir, the normalization in
Eq.~\eqref{eq:II_harmonic_definition} gives the measured linear change
\begin{equation}
 \delta I_b^{(3),\mathrm{AB}}(\Phi)
 =\frac{2q|\Gamma_{02}\Gamma_{21}\Gamma_{10}|}{\hbar^3}
 \operatorname{Re}\!\left[e^{\mathrm{i}\varphi_{\mathrm{AB}}}
                        \delta\mathcal I_b\right],
 \label{eq:IV_bridge_observable}
\end{equation}
where $\varphi_{\mathrm{AB}}$ is the Laughlin specialization of
Eq.~\eqref{eq:II_loop_phase}, and
$\delta\mathcal I_b=\lambda_a(\partial_{\lambda_a}\mathcal I_b)_0$
includes the bridge coupling.  It is this terminal coefficient, rather
than an isolated response between two points, that is calculated below.

The interaction acts only within $C_a$:
\begin{align}
 H_{U,a}&=\frac{\lambda_a}{2}\int dy\,dz\,
          U_a(y,z)n_a(y)n_a(z),
 \nonumber\\
 n_a(x)&=\frac{\partial_x\Phi_a(x)}{2\pi},\qquad
 \Phi_a=\sqrt\nu\,\varphi_a.
 \label{eq:bridge_HU}
\end{align}
Here $\varphi_a$ is the normalized chiral field, and $\Phi_a$ is its
quasiparticle vertex phase, not the magnetic flux $\Phi$.  The small
parameter $\lambda_a$ is dimensionless; the real, instantaneous kernel
$U_a(y,z)=U_a(z,y)$ has units of energy and is finite-range or sufficiently
rapidly decaying.  The complete quadratic Hamiltonian is stable.  The density
is a fluctuation about its reservoir-fixed mean, with static mean shifts
included in the calibrated voltages and common QPC phase.  There is no direct
quadratic coupling between distinct reservoir-fed sectors.

The relevant spatial connection is shown in Fig.~\ref{fig:static_bridge}:
\begin{equation}
 y<x_{a,u}<x_{a,d}<z.
 \label{eq:bridge_cross_support}
\end{equation}
A perturbation at $x_{a,d}$ propagates downstream to $z$.  The density
coupling links $z$ to $y$, from which the response propagates downstream to
$x_{a,u}$.  Thus both propagation legs have positive velocity $v_a$, while
the nonlocal interaction permits the downstream QPC to influence the upstream
one.  The interaction transfers no charge between terminal sectors:
$[Q_b,H_{U,a}]=0$.

\subsection{Interaction-induced response and static limit}

The interaction modifies the existing edge correlations and can also generate
the reverse response illustrated in Fig.~\ref{fig:static_bridge}.  To compute
the complete first-order correction, we include its effect on both propagation
and the stationary equilibrium state in the nested-commutator current,
retaining both time orderings and both loop orientations.

At equal drain voltages, the contribution from changes in the existing edge
correlations enters the two adjacent QPC current terms with the same
detailed-balance factors and cancels in their difference.  This cancellation
requires a common temperature, the same quasiparticle exchange phase, and
independent stationary sectors, but not identical drain spectra or
normalizations.  The remaining dark signal depends on the nonlocal response
from the downstream QPC to the upstream QPC.  Its contribution to the measured
current nevertheless contains both time orderings of
Sec.~\ref{sec:cubic_current}.  Appendix~\ref{app:bridge} derives the
cancellation and combines all remaining terms into the complete linear
current.

When bias and temperature do not resolve the duration of this response, its
time dependence enters only through an integrated strength, expressed below
as the spatial moment $\mathcal Q_{\nu,a}[U_a]$.  The other two edges supply
the spectral weight for transfer at energy $E$, with the reverse transfer
subtracted by detailed balance.  We denote this net two-edge spectrum by
$P_\nu^{\rm net}(E,T)$.  The dark harmonic therefore factorizes into the
spatial interaction strength and the bias-dependent spectrum of the other
two edges.  For the common local Laughlin fixed point, the result is
\begin{equation}
 \begin{aligned}
 \delta\mathcal I_{a-1}(E,T)&\simeq
 -\frac{\lambda_a\nu^2}{\hbar v_a^2}\mathcal Q_{\nu,a}[U_a]\\
 &\quad\times e^{\mathrm{i}\pi\nu/2}P_\nu^{\rm net}(E,T).
 \end{aligned}
 \label{eq:bridge_dark_master}
\end{equation}
For $E>0$ and a nonzero leading coefficient, the response phase in
Eq.~\eqref{eq:II_amplitude_phase} is
$\pi\nu/2+\arg[-\lambda_a\mathcal Q_{\nu,a}]$ modulo $2\pi$.
The second term is fixed by the sign of the real prefactor.
The real signed spatial moment is
\begin{equation}
 \mathcal Q_{\nu,a}[U_a]
 =\int_{-\infty}^{x_{a,u}}dy\int_{x_{a,d}}^\infty dz\,
 U_a(y,z)\left(\frac{v_a\tau_0}{z-y}\right)^\nu.
 \label{eq:bridge_Qmoment}
\end{equation}
The operator-normalization time $\tau_0$ is the one defined in
Appendix~\ref{app:vertex_loop}; it is distinct from the physical ultraviolet
scale $E_{\rm uv}$.  The fractional factor contains the full separation
$z-y$, not the excess distance outside the QPC interval.

The spectrum of the other two edges, evaluated within the scaling theory, is
\begin{align}
 P_\nu(E,T)&=\frac{\tau_0^{2\nu}}{\Gamma(2\nu)}
       \left(\frac{2\pi T}{\hbar}\right)^{2\nu-1}
       e^{E/(2T)}
 \nonumber\\[-1mm]
 &\quad\times
       \left|\Gamma\!\left(\nu+\frac{\mathrm{i}E}{2\pi T}\right)\right|^2,
 \nonumber\\
 P_\nu^{\rm net}(E,T)&=P_\nu(E,T)-P_\nu(-E,T)
 \nonumber\\
 &=(1-e^{-E/T})P_\nu(E,T).
 \label{eq:bridge_Pnet}
\end{align}
It is positive for $E>0$ and vanishes at $E=0$, so the interaction-induced
current also vanishes in full equilibrium.  This positive spectral weight
does not make the AB interference contribution a positive transfer rate.
Since $\mathcal Q_{\nu,a}$ has units of energy times length squared, the
right-hand side of Eq.~\eqref{eq:bridge_dark_master} has units of time
squared, as required for the reduced terminal coefficient.

\subsection{Scaling and the controlled operating window}

Since the reference harmonic at this drain is zero, the magnitude of the
linear correction gives the leading activated amplitude.  Applying
Eq.~\eqref{eq:II_amplitude_phase} to $\delta\mathcal I_{a-1}$ gives
\begin{equation}
 A_{a-1}^{\mathrm{dark}}
 =\frac{2q|\Gamma_{02}\Gamma_{21}\Gamma_{10}|}{\hbar^3}
       |\delta\mathcal I_{a-1}|.
 \label{eq:bridge_physical_amplitude}
\end{equation}
At fixed bridge profile and QPC amplitudes, the high-bias expansion is
\begin{equation}
 P_\nu^{\rm net}(E,T)
 =\frac{2\pi\tau_0^{2\nu}}{\Gamma(2\nu)}
 \left(\frac E\hbar\right)^{2\nu-1}
 \left[1+O\!\left((T/E)^2\right)\right].
 \label{eq:bridge_Pasymptotic}
\end{equation}
Consequently, at leading static order and away from a zero of the moment,
\begin{equation}
 A_{a-1}^{\mathrm{dark}}\propto E^{2\nu-1}.
 \label{eq:bridge_dark_scaling}
\end{equation}
The spectra of the other two edges determine this power: reducing the
intermediate response to a finite moment removes its long-time integration
from the bias scaling.

The high-bias measurement requires
\begin{equation}
 T_{B,0}\ll T\ll E\ll E_{\rm uv},
 \label{eq:IV_bridge_window}
\end{equation}
where $T_{B,0}$ is the strong-coupling crossover scale of the unbiased QPC.
That QPC must remain weak at energy $T$; the other two must remain weak at
their source--drain scales.  The edge-flight inequalities in
Eq.~\eqref{eq:II_unresolved_flight} also apply.  Finite temperature is thus
an essential infrared cutoff, not a parameter that can be taken to zero at
fixed relevant tunneling strength.

For the static reduction, the extra response time of a cross-region pair is
$(z-y-D_a)/v_a$.  On appreciable support one requires
\begin{equation}
 \frac{\max(E,T)(z-y-D_a)}{\hbar v_a}\ll1,
 \qquad \frac{T(z-y)}{\hbar v_a}\ll1.
 \label{eq:bridge_unresolved_condition}
\end{equation}
The first inequality permits the spectrum of the other two edges to be
evaluated at zero extra response time; the second
permits the thermal spatial moment to be replaced by
Eq.~\eqref{eq:bridge_Qmoment}.  The contributing distances must also lie
within the physical edge-scaling regime.

A local kernel, or one with no cross-region support, gives no dark activation
at first bridge order.  A spanning kernel can nevertheless have
$\mathcal Q_{\nu,a}=0$ by cancellation.  Such a zero removes only the leading
static term: higher response moments may control the residual signal and its
phase.  Small geometric delays alone do not control a relative error near
that zero.

\subsection{Bright reference and interpretation}

The bridge-free bright harmonic retains a long correlation time set by the
unbiased QPC.  For $0<\nu<1/2$ and $T\ll E$, one time integral is dominated
by times of order $\hbar/E$, while the other is cut off at $\hbar/T$.
Their leading factors are $E^{\nu-1}$ and $T^{2\nu-1}$, respectively.
Thus, in the window of Eq.~\eqref{eq:IV_bridge_window},
\begin{equation}
 |F_\nu^{\mathrm d}(E,0;T)|\propto E^{\nu-1}T^{2\nu-1}.
 \label{eq:IV_bridge_bright_asymptotic}
\end{equation}
The endpoint estimate and its normalization are given in
Appendix~\ref{app:C_reference}.  The function $F_\nu^{\mathrm d}$ belongs
to the local theory, but a weak, localized finite-range interaction in the
regular single-mode infrared regime preserves its leading bias and
temperature powers at the order considered.  The complete Gaussian
calculation in Eq.~\eqref{eq:C_bright_renormalized} shows that the bright
prefactor is renormalized, while the leading powers remain those of
Eq.~\eqref{eq:IV_bridge_bright_asymptotic}.  The measured bright harmonic can
therefore serve as a scaling reference in the same device without switching
off the natural interaction, provided the relevant interaction scales are
also unresolved.
The equal-drain power differs both from the activated law
\eqref{eq:bridge_dark_scaling} and from the fixed-nonzero-detuning local
homogeneity $E^{3\nu-2}$.  The unbiased QPC makes the equal-drain reference
infrared sensitive.  At $\nu=1/3$, the leading dark contribution and the
bridge-free bright reference scale as $E^{-1/3}$ and
$E^{-2/3}T^{-1/3}$, respectively.  The exact thermal expression, not a
high-bias asymptotic, gives the equilibrium zero.

The interaction in Eq.~\eqref{eq:bridge_HU} can describe a weak natural
nonlocal coupling; no sign-reversal device is needed to define this model or
its calculated activation.  The result shows that a dark-terminal signal
need not imply intrinsic upstream content of the selected excitation.
The support and scaling predictions are diagnostics within the stated
single-Laughlin-bridge model, not a model-independent decomposition of a
signal containing intrinsic upstream weight, several interactions or
higher-QPC-order backgrounds.  In particular, the local amplitude-ratio
inversion of Sec.~\ref{sec:kmatrix} must not be applied to an arbitrary
interacting ratio without controlling the nonlocal contribution.

The measurement implications and the limits of this comparison are discussed
in Sec.~\ref{sec:discussion}.

\section{Discussion and conclusions}
\label{sec:discussion}
\label{sec:conclusion}

The three-path interferometer makes the directional content of a tunneling
excitation accessible through an average electrical current.  The relevant
distinction is between the propagation of electric charge and the propagation
of the correlations that govern tunneling.  In the regime considered here,
injected charge is collected downstream, but an upstream neutral mode can
still affect interference when it participates in the selected quasiparticle
operator.  The measurement therefore probes the directional structure of that
excitation, rather than simply the direction of electrical transport.

The simplest manifestation is the equal-drain null in
Eq.~\eqref{eq:K_equal_drain_null}.  At one drain, charge balance cancels the
downstream contribution to the cubic fundamental AB harmonic.  If the selected
excitation has no upstream component, the remaining contribution also vanishes.
The neighboring drain retains an oscillation, while the device remains biased
and all three QPC amplitudes are nonzero.  This is the absence of an entire
harmonic at one terminal, not a minimum reached by tuning the flux.  It turns
the directional restriction of the local theory into a directly identifiable
feature of the current oscillations.

% The old measurement-subsection label now identifies this paragraph in Sec. VI.
\phantomsection
\label{subsec:protocol_window}
Exchanging two unequal drain voltages extends this separation to a regime in
which every QPC remains biased.  After compensating the independently
calibrated electrostatic phase shift, averaging the two flux traces isolates
the upstream response at one drain and the downstream response at the other.
The QPC magnitudes must remain fixed or independently known, both between the
paired settings and along the subsequent bias sweep.  The amplitudes used for
spectroscopy are those of these averaged oscillations.

The two amplitudes contain complementary information.  At fixed nonzero
drain-detuning ratio and temperature small compared with every QPC bias
energy, their common power $E^{3\Delta_{\bm\ell}-2}$ determines the scaling
dimension $h_{\bm\ell}=\Delta_{\bm\ell}/2$.  Their ratio then determines how
the total vertex weight is distributed between downstream and upstream modes.
Together, these measurements determine the principal exchange angle for
$0<\Delta_{\bm\ell}<1$, through
Eqs.~\eqref{eq:K_directional_ratio}--\eqref{eq:K_weight_reconstruction}.
The quasiparticle charge remains an independently calibrated input, obtainable,
for example, from weak-QPC shot noise~\cite{Saminadayar1997,dePicciotto1997}.

The distinction between scaling and directional information is especially
clear at the disorder-dominated $\nu=2/3$ fixed point: a selected charge-$e/3$
excitation and the neutral-free charge-$2e/3$ excitation have the same scaling
dimension, but directional amplitude ratios of $2$ and $0$, respectively.
The latter does not couple to the upstream neutral mode even though that mode
remains present on the edge.

The nonlocal-interaction calculation gives a complementary result.  A density
coupling between points on opposite sides of the QPC pair can join two
downstream propagation processes and allow the downstream contact to influence
the upstream contact.  Thus an otherwise dark harmonic can appear without an
upstream propagating mode.  For the weak Laughlin interaction studied in
Sec.~\ref{sec:bridge}, the leading static response separates into a spatial
interaction moment and the net tunneling spectrum of the other two edges.
For a nonzero moment, its high-bias amplitude scales as $E^{2\nu-1}$; the local
bright reference instead scales as $E^{\nu-1}T^{2\nu-1}$ for $0<\nu<1/2$ in
the stated thermal window.  These different powers connect the spatial
mechanism to a measurable bias dependence.  For a weak, localized finite-range
coupling in the regular single-mode infrared regime, the bright harmonic
retains these leading bias and temperature powers at the interaction order
considered, although its prefactor is renormalized.  It can therefore serve as
a scaling reference in the same device, without requiring the natural
interaction to be switched off.

% The old interpretation-subsection label now identifies this paragraph in Sec. VI.
\phantomsection
\label{subsec:protocol_interpretation}
The results apply when weak tunneling of one coherent quasiparticle species
dominates and the stationary charge and neutral correlations satisfy the
common-temperature and unresolved-flight conditions stated in
Sec.~\ref{sec:setup}.  Within the local regime, a directional signal below
experimental sensitivity still yields useful information: its upper bound,
combined with the resolved bright amplitude and measured exponent, constrains
the corresponding directional weight.  The intrinsic reconstruction requires
residual calibration errors and additional coherent contributions, including
nonlocal interactions, to be negligible or independently characterized.
Higher tunneling orders and competing species are outside this leading
single-vertex interpretation.

Together, the null measurement, directional amplitude ratio, and bias scaling
provide access to complementary properties of the selected tunneling
excitation using flux-dependent average currents.  The interaction result
identifies spatial locality as a separate physical ingredient in this
connection.  The three-path geometry thereby relates an electrical
interference signal to both the directional correlations of the tunneling
excitation and its scaling and statistical properties.

\section*{Acknowledgments}
We thank Ivan Levkivskyi for fruitful discussions.

\bibliographystyle{apsrev4-2}
\bibliography{References}

\appendix

\section{Nested-commutator derivation of the cubic terminal current}
\label{app:cubic_current}

This appendix derives the terminal-current expressions
\eqref{eq:III_terminal_d} and \eqref{eq:III_terminal_u} from the current
operator \eqref{eq:II_current_operator}, in the voltage-phase representation
of Sec.~\ref{sec:setup}.  We obtain the cubic real-time expansion, identify
its two causal contributions, and combine the two QPC terms associated with
each terminal.  Appendix~\ref{app:vertex_loop} evaluates the vertex averages.

\subsection{Current operator and real-time expansion}
\label{app:A_current_expansion}

The operator $\mathcal A_{a,a-1}$ transfers one quasiparticle of charge
$q_{\boldsymbol\ell}$ from sector $a-1$ to sector $a$.  Therefore
\begin{equation}
 [Q_b,\mathcal A_{a,a-1}]
 =q_{\boldsymbol\ell}(\delta_{ba}-\delta_{b,a-1})
 \mathcal A_{a,a-1}.
 \label{eq:A_charge_commutator}
\end{equation}
Since $[Q_b,K_0]=0$, the tunneling current, positive into the collecting
reservoir as in Eq.~\eqref{eq:II_terminal_current_definition}, is
\begin{align}
 \widehat I_b(t)
 &=\left.\dot Q_b(t)\right|_T
   =\frac{\mathrm{i}}{\hbar}[H_T(t),Q_b]
 \nonumber\\[-1mm]
 &=-\frac{\mathrm{i}q_{\boldsymbol\ell}}{\hbar}
 \sum_a(\delta_{ba}-\delta_{b,a-1})
 \left[
 \mathcal T_{a,a-1}(t)-\mathcal T_{a,a-1}^{\dagger}(t)
 \right].
 \label{eq:A_current_operator}
\end{align}
The opposite signs at the ends of each link give
$\sum_b\widehat I_b=0$.

The exact expectation value is
\begin{equation}
 \langle I_b(0)\rangle
 =\operatorname{Tr}\!\left[
 \rho_0U^\dagger(0,t_0)\widehat I_b(0)U(0,t_0)
 \right],
 \label{eq:A_closed_contour}
\end{equation}
where $U(0,t_0)=T\exp[-(\mathrm{i}/\hbar)\int_{t_0}^{0}dt\,H_T(t)]$
is the time-ordered evolution operator.  Unitarity gives
$\operatorname{Tr}(\rho_0U^\dagger U)=1$, so no separate normalization
factor is required.  Since $\widehat I_b$ is already first order in the QPC
amplitudes, the cubic current follows by expanding $U$ and $U^\dagger$
through second order.  Splitting the cross term into its two time-ordering
regions and collecting all terms over $t_2<t_1<0$ gives
\begin{align}
 &\widehat I_bH_T(t_1)H_T(t_2)
 -H_T(t_2)\widehat I_bH_T(t_1)
 \nonumber\\[-1mm]
 &\hspace{12mm}
 -H_T(t_1)\widehat I_bH_T(t_2)
 +H_T(t_2)H_T(t_1)\widehat I_b
 \nonumber\\[-1mm]
 &\hspace{34mm}
 =[[\widehat I_b,H_T(t_1)],H_T(t_2)].
 \label{eq:A_branch_sum}
\end{align}
Here the current operator is evaluated at $t=0$.  With adiabatic switch-on
and $t_0\to-\infty$, the result is
\begin{align}
 \langle I_b(0)\rangle^{(3)}
 ={}&-\frac{1}{\hbar^2}
 \int_{-\infty}^{0}dt_1
 \int_{-\infty}^{t_1}dt_2
 \nonumber\\[-1mm]
 &\times\left\langle
 [[\widehat I_b(0),H_T(t_1)],H_T(t_2)]
 \right\rangle_0.
 \label{eq:A_nested_current}
\end{align}
This is Eq.~\eqref{eq:II_nested_commutator}, obtained directly from the
normalized unitary evolution.

\subsection{The two directional loop contributions}
\label{app:A_directional_loops}

The positive fundamental harmonic contains one directed operator on each
QPC.  Consider the contribution of the QPC $a-1\to a$ to the current
operator at $t=0$.  The two remaining QPC operators act at earlier times,
$t_2<t_1<0$, and define
\begin{align}
 M_a^{\mathrm d}(t_1,t_2)
 &=\left\langle
 [[\mathcal A_{a,a-1}(0),\mathcal A_{a-1,a+1}(t_1)],
   \mathcal A_{a+1,a}(t_2)]
 \right\rangle_0,
 \label{eq:A_Md}\\
 M_a^{\mathrm u}(t_1,t_2)
 &=\left\langle
 [[\mathcal A_{a,a-1}(0),\mathcal A_{a+1,a}(t_1)],
   \mathcal A_{a-1,a+1}(t_2)]
 \right\rangle_0.
 \label{eq:A_Mu}
\end{align}
The first inner commutator shares $C_{a-1}$ and connects its upstream QPC at
time $t_1$ to its downstream QPC at the later time $0$.  The second shares
$C_a$ and connects its downstream QPC at $t_1$ to its upstream QPC at $0$.
For a fixed segment $C_a$, the two directions are therefore represented by
$M_{a+1}^{\mathrm d}$ and $M_a^{\mathrm u}$, respectively, as illustrated in
Fig.~\ref{fig:causal_paths}.

Using the QPC phases $\Theta_a$ defined in Eq.~\eqref{eq:III_Theta} and the
fixed cross-edge ordering of Appendix~\ref{app:B_modes_ordering}, the mode
algebra gives, for $t>t'$,
\begin{align}
 [\Theta_a(t),\Theta_{a-1}(t')]
 &=2\pi\mathrm{i}\sum_{v_j>0}p_j^2
 \Theta_{\rm H}\!\left(t-t'-\frac{D_{a-1}}{|v_j|}\right),
 \label{eq:A_down_support}\\
 [\Theta_a(t),\Theta_{a+1}(t')]
 &=2\pi\mathrm{i}\sum_{v_j<0}p_j^2
 \Theta_{\rm H}\!\left(t-t'-\frac{D_a}{|v_j|}\right).
 \label{eq:A_up_support}
\end{align}
These step functions retain the propagation thresholds before the short
delays are neglected under Eq.~\eqref{eq:II_unresolved_flight}.  For
$\delta_-=0$, the upstream inner QPC commutator vanishes as an operator,
so $M_a^{\mathrm u}(t_1,t_2)=0$ for every ordered pair of times, before any
time integration.

Restoring the voltage phases, define the reduced positive-loop contributions
associated with QPC $a-1\to a$ by
\begin{align}
 \mathcal J_a^{\mathrm d}
 ={}&-\mathrm{i}\int_{-\infty}^{0}dt_1
 \int_{-\infty}^{t_1}dt_2\,
 \nonumber\\[-1mm]
 &\times
 e^{-\mathrm{i}[\epsilon_{a+1,a-1}t_1+
                  \epsilon_{a,a+1}t_2]/\hbar}
 M_a^{\mathrm d}(t_1,t_2),
 \label{eq:A_Jd}\\
 \mathcal J_a^{\mathrm u}
 ={}&-\mathrm{i}\int_{-\infty}^{0}dt_1
 \int_{-\infty}^{t_1}dt_2\,
 \nonumber\\[-1mm]
 &\times
 e^{-\mathrm{i}[\epsilon_{a,a+1}t_1+
                  \epsilon_{a+1,a-1}t_2]/\hbar}
 M_a^{\mathrm u}(t_1,t_2).
 \label{eq:A_Ju}
\end{align}
These auxiliary QPC-resolved integrals are used only to construct the
terminal current.  Their differences are formed before removing the common
short-time regulator, as discussed in Appendix~\ref{app:vertex_loop}.

\subsection{Terminal combination and perturbative order}
\label{app:A_terminal_combination}

Substituting Eq.~\eqref{eq:A_current_operator} into
Eq.~\eqref{eq:A_nested_current} and factoring out the positive loop product
$\Gamma_{\circlearrowright}$ gives
\begin{align}
 \mathcal I_b^{\eta}
 &=\mathcal J_{b+1}^{\eta}-\mathcal J_b^{\eta},
 \qquad \eta=\mathrm d,\mathrm u,
 \label{eq:A_terminal_components}\\
 \mathcal I_b
 &=\mathcal I_b^{\mathrm d}+\mathcal I_b^{\mathrm u}
 =\mathcal J_{b+1}-\mathcal J_b,
 \qquad
 \mathcal J_a=\mathcal J_a^{\mathrm d}+\mathcal J_a^{\mathrm u}.
 \label{eq:A_terminal_incidence}
\end{align}
Both QPC terms adjacent to terminal $b$ are included.  Their relative sign
is fixed by charge injection in Eq.~\eqref{eq:A_current_operator}; the
$\mathcal J_a^{\eta}$ are not separate terminal currents.  Adding the
Hermitian-conjugate loop gives the real harmonic in
Eq.~\eqref{eq:II_harmonic_definition}.

Set $u=-t_1>0$ and $v=t_1-t_2>0$.  In the local unresolved-flight regime,
Appendix~\ref{app:vertex_loop} gives
\begin{equation}
 \begin{aligned}
 M_a^{\mathrm d}(t_1,t_2)
 &=\alpha_{\mathrm d}M_{\Delta_{\boldsymbol\ell}}^{(0)}(u,v;T),\\
 M_a^{\mathrm u}(t_1,t_2)
 &=e^{\mathrm{i}\theta_{\boldsymbol\ell}}\alpha_{\mathrm u}
 M_{\Delta_{\boldsymbol\ell}}^{(0)}(u,v;T).
 \end{aligned}
 \label{eq:A_factorization}
\end{equation}
Here $-\mathrm{i}M_{\Delta_{\boldsymbol\ell}}^{(0)}=
A_{\Delta_{\boldsymbol\ell}}W_{\Delta_{\boldsymbol\ell}}$.
The explicit upstream factor is the relative phase of the ordinary ordered
products; it is not included in the exchange-coefficient ratio
$\alpha_{\mathrm u}$.

The cyclic drops of Eq.~\eqref{eq:III_cyclic_drops} obey
$\epsilon_{a+1,a-1}+\epsilon_{a,a+1}=-\varepsilon_{a-1}$ and
$\epsilon_{a+1,a-1}=\varepsilon_{a+1}$.  The voltage factors in
Eqs.~\eqref{eq:A_Jd} and \eqref{eq:A_Ju} therefore become
$e^{-\mathrm{i}\varepsilon_{a-1}u/\hbar+
\mathrm{i}\varepsilon_av/\hbar}$ and
$e^{-\mathrm{i}\varepsilon_{a-1}u/\hbar+
\mathrm{i}\varepsilon_{a+1}v/\hbar}$, respectively.  Combining them as
$\mathcal J_{b+1}^{\eta}-\mathcal J_b^{\eta}$ produces precisely the
ordered differences of exponentials in Eqs.~\eqref{eq:III_terminal_d} and
\eqref{eq:III_terminal_u}.  The two terms in each bracket thus come from
the two QPC contributions to the terminal-current operator.

Summing the terminal differences gives zero, as in
Eq.~\eqref{eq:III_conservation}.  At equal reservoir voltages, the
voltage-dependent differences in the local terminal integrals vanish.
The regulated free-electron limit is evaluated in
Appendix~\ref{app:B_contact_limit} and reproduces the Landauer current.

Finally, invariance under independent rephasing of the three edge sectors
requires each positive fundamental-harmonic monomial to contain one excess
directed amplitude on every link.  It therefore contains
$\Gamma_{02}\Gamma_{21}\Gamma_{10}$, with any additional amplitudes entering
in conjugate pairs.  The allowed QPC orders are
\begin{equation}
 N=3+2p,\qquad p=0,1,2,\ldots .
 \label{eq:A_order_count}
\end{equation}
There is no fourth-order fundamental harmonic; the first QPC-network
correction to the cubic null is allowed at fifth order.

\section{Abelian vertex algebra and terminal loop kernels}
\label{app:vertex_loop}

We evaluate the vertex averages entering the terminal current derived in
Appendix~\ref{app:cubic_current}.  The exchange algebra determines the
directional factors, while the local correlators supply a common thermal
loop function.  We then establish the convergence and nonvanishing needed
for the spectroscopy in Sec.~\ref{sec:kmatrix}, and give the regulated
free-electron limit used in Sec.~\ref{subsec:K_equal_drains}.  Throughout
this appendix, $\Delta\equiv\Delta_{\boldsymbol\ell}$.

\subsection{Vertex correlations and directional factors}
\label{app:B_modes_ordering}
\label{app:B_exchange_algebra}
\label{app:B_correlator_loop}

The propagation basis of Sec.~\ref{subsec:abelian_model} is obtained from
\begin{align}
 \mathsf V_{\rm edge}\,\boldsymbol u_j&=v_jK\boldsymbol u_j,
 \label{eq:B_generalized_eigenproblem}\\
 \boldsymbol u_j^TK\boldsymbol u_k&=\sigma_j\delta_{jk},
 \qquad \sigma_j=\operatorname{sgn}v_j.
 \label{eq:B_mode_normalization}
\end{align}
Positivity of $\mathsf V_{\rm edge}$ gives real velocities and permits this
normalization.  Taking the eigenvectors as columns of $M$ gives
Eq.~\eqref{eq:II_mode_normalization}, with $u_j=|v_j|$ and
$\boldsymbol p=M^T\boldsymbol\ell$.  The identities
$K^{-1}=M\Sigma M^T$ and $\Delta=\sum_jp_j^2$ then give the sum and
difference of directional weights stated in
Eqs.~\eqref{eq:II_directional_weights} and \eqref{eq:II_sum_difference}.

For operators on distinct reservoir-fed sectors, we retain one fixed global
order by connecting the sectors at infinity along an auxiliary contour:
\begin{equation}
 s_{0,u}<s_{0,d}<s_{1,u}<s_{1,d}<s_{2,u}<s_{2,d}<s_{0,u}+L_s.
 \label{eq:B_superedge_order}
\end{equation}
Here $s$ is the contour coordinate and $L_s$ its total length.  The QPC
bilinears occupy adjacent, noncrossing intervals, so their equal-time
exchange phases cancel.  This convention is equivalent to explicit
cocycles or Klein factors~\cite{Guyon2002,vonDelftSchoeller1998}, but requires
no independent Klein-factor dynamics for the neutral cubic loop.  Common
constant phases are included in the QPC-product phase of
Eq.~\eqref{eq:II_loop_phase}.  With $\mathcal V^+=\mathcal V$ and
$\mathcal V^-=\mathcal V^\dagger$, the exchange relation is
\begin{align}
 \mathcal V^\sigma(s)\mathcal V^{\sigma'}(s')
 ={}&e^{-\mathrm{i}\sigma\sigma'\theta_{\boldsymbol\ell}
                \operatorname{sgn}(s-s')}
 \nonumber\\[-1mm]
 &\times\mathcal V^{\sigma'}(s')\mathcal V^\sigma(s).
 \label{eq:B_global_vertex_exchange}
\end{align}

In the local, unresolved-flight regime, the temporal vertex correlators are
\begin{align}
 g_\Delta^>(t)
 &=\lim_{\epsilon\to0^+}
 \left[\frac{\pi T\tau_0/\hbar}
 {\sin\{\pi T(\epsilon+\mathrm{i}t)/\hbar\}}\right]^\Delta,
 \label{eq:B_greater}\\
 g_\Delta^<(t)&=g_\Delta^>(-t).
 \label{eq:B_lesser}
\end{align}
Here $T$ is in energy units, $\epsilon$ specifies the fractional-power
branch, and $\tau_0$ fixes the operator normalization.  The physical
upper validity scale $E_{\rm uv}$ is separate; no relation
$E_{\rm uv}=\hbar/\tau_0$ is assumed.  Short-distance normalization is
absorbed into the QPC amplitudes as in Sec.~\ref{subsec:tunneling_loop}.
For $t>0$, their standard scaling form
\cite{Wen1990,ChamonFreedWen1996,FendleyLudwigSaleur1995,Chang2003} is
\begin{align}
 g_\Delta^>(t)&=g_\Delta^<(-t)
   =e^{-\mathrm{i}\pi\Delta/2}\mathcal E_\Delta(t,T),
 \nonumber\\[-1mm]
 \mathcal E_\Delta(t,T)
 &=\left[\frac{\pi T\tau_0/\hbar}{\sinh(\pi Tt/\hbar)}\right]^\Delta.
 \label{eq:B_thermal_envelope}
\end{align}
The envelope $\mathcal E_\Delta$ is real and positive; the analytic phase
must be retained separately.

To evaluate the nested commutators, let $XY=q_{XY}YX$,
$XZ=q_{XZ}ZX$, and $YZ=q_{YZ}ZY$, with c-number exchange factors.
Expanding and reordering gives
\begin{align}
 [[X,Y],Z]&=XYZ-YXZ-ZXY+ZYX
 \nonumber\\[-1mm]
 &=(1-q_{XY}^{-1})(1-q_{XZ}^{-1}q_{YZ}^{-1})XYZ.
 \label{eq:B_nested_exchange_identity}
\end{align}
For the two orderings in Eqs.~\eqref{eq:A_Md} and \eqref{eq:A_Mu}, $X$
is the QPC operator at $0$, while $Y_\eta$ and $Z_\eta$ are those at
$t_1$ and $t_2$, respectively, with $t_2<t_1<0$.  The support formulas
\eqref{eq:A_down_support} and \eqref{eq:A_up_support} give
\begin{equation}
 \renewcommand{\arraystretch}{1.15}
 \begin{array}{c|ccc}
 \eta & q_{XY_\eta}^{-1} & q_{XZ_\eta}^{-1} & q_{Y_\eta Z_\eta}^{-1}\\ \hline
 \mathrm d & e^{2\pi\mathrm{i}\delta_+}
           & e^{2\pi\mathrm{i}\delta_-} & e^{2\pi\mathrm{i}\delta_+}\\
 \mathrm u & e^{2\pi\mathrm{i}\delta_-}
           & e^{2\pi\mathrm{i}\delta_+} & e^{2\pi\mathrm{i}\delta_-}
 \end{array}
 \label{eq:B_pair_exchange_factors}
\end{equation}
The spatial and temporal order is fixed before neglecting the flight
delays.  The second factor in Eq.~\eqref{eq:B_nested_exchange_identity}
is $1-e^{2\pi\mathrm{i}\Delta}$ in either row.  Relative to the downstream
reference with $(\delta_+,\delta_-)=(\Delta,0)$, whose coefficient is
$(1-e^{2\pi\mathrm{i}\Delta})^2$, the directional factors are therefore
\begin{align}
 \alpha_{\mathrm d}
 &=\frac{1-e^{2\pi\mathrm{i}\delta_+}}{1-e^{2\pi\mathrm{i}\Delta}}
 =e^{-\mathrm{i}\pi\delta_-}\frac{\sin(\pi\delta_+)}{\sin(\pi\Delta)},
 \label{eq:B_alpha_d}\\
 \alpha_{\mathrm u}
 &=\frac{1-e^{2\pi\mathrm{i}\delta_-}}{1-e^{2\pi\mathrm{i}\Delta}}
 =e^{-\mathrm{i}\pi\delta_+}\frac{\sin(\pi\delta_-)}{\sin(\pi\Delta)}.
 \label{eq:B_alpha_u}
\end{align}
These are the ratios in Eqs.~\eqref{eq:III_alpha_d} and
\eqref{eq:III_alpha_u}.  They normalize the exchange coefficients, not the
ordinary products multiplying them.  At an integer contact point the
complete regulated expression is used instead of these separate ratios.

The remaining averages contain one vertex pair on each edge.  Put
$u=-t_1>0$ and $v=t_1-t_2>0$.  Regrouping the six vertices by edge,
without reversing either pair on the same edge, gives
\begin{align}
 \langle XY_{\mathrm d}Z_{\mathrm d}\rangle_0
 &=g_\Delta^>(u)g_\Delta^>(v)g_\Delta^<(-u-v),
 \label{eq:B_down_ordered_product}\\
 \langle XY_{\mathrm u}Z_{\mathrm u}\rangle_0
 &=e^{\mathrm{i}\theta_{\boldsymbol\ell}}
   g_\Delta^<(-u)g_\Delta^<(-v)g_\Delta^>(u+v).
 \label{eq:B_up_ordered_product}
\end{align}
The downstream regrouping has unit phase in the order
\eqref{eq:B_superedge_order}.  For example, for the QPC $0\to1$, the
upstream sequence is
$(1u^\dagger,0d,2u^\dagger,1d,0u^\dagger,2d)$, where each entry denotes
the vertex at that endpoint.  The cross-edge exchanges in
Eq.~\eqref{eq:B_global_vertex_exchange} contribute a net
$+\theta_{\boldsymbol\ell}$.  Thus the two products share a real envelope,
but retain a relative phase:
\begin{equation}
 \begin{aligned}
 \langle XY_{\mathrm d}Z_{\mathrm d}\rangle_0
 &=e^{-3\mathrm{i}\pi\Delta/2}W_\Delta(u,v;T),\\
 \langle XY_{\mathrm u}Z_{\mathrm u}\rangle_0
 &=e^{\mathrm{i}\theta_{\boldsymbol\ell}-3\mathrm{i}\pi\Delta/2}
   W_\Delta(u,v;T),
 \end{aligned}
 \label{eq:B_common_ordered_product}
\end{equation}
where
\begin{equation}
 W_\Delta(u,v;T)
 =\mathcal E_\Delta(u,T)\mathcal E_\Delta(v,T)\mathcal E_\Delta(u+v,T).
 \label{eq:B_W_definition}
\end{equation}
Multiplying by the reference exchange coefficient gives
\begin{equation}
 M_\Delta^{(0)}(u,v;T)
 =-4e^{\mathrm{i}\pi\Delta/2}\sin^2(\pi\Delta)W_\Delta(u,v;T).
 \label{eq:B_reference_nested}
\end{equation}
The complete factorization is
\begin{align}
 M_a^{\mathrm d}&=\alpha_{\mathrm d}M_\Delta^{(0)},
 &M_a^{\mathrm u}&=e^{\mathrm{i}\theta_{\boldsymbol\ell}}
                    \alpha_{\mathrm u}M_\Delta^{(0)},
 \nonumber\\[-1mm]
 -\mathrm{i}M_\Delta^{(0)}&=A_\Delta W_\Delta,
 &A_\Delta&=4\mathrm{i}e^{\mathrm{i}\pi\Delta/2}\sin^2(\pi\Delta).
 \label{eq:B_factorization}
\end{align}
Inserting this result in Eqs.~\eqref{eq:A_Jd} and \eqref{eq:A_Ju}, and
forming the collected-current difference
$\mathcal J_{b+1}^\eta-\mathcal J_b^\eta$ of
Eq.~\eqref{eq:A_terminal_components}, gives
Eqs.~\eqref{eq:III_terminal_d} and \eqref{eq:III_terminal_u}.

\subsection{Scalar response and its domain of validity}
\label{app:B_uv_domain}
\label{app:B_scalar_support}

The terminal difference is essential for the short-time behavior.  Denote
the brackets already displayed in Eqs.~\eqref{eq:III_terminal_d} and
\eqref{eq:III_terminal_u} by $B_b^\eta(u,v)$, with
$\eta=\mathrm d,\mathrm u$.  Both vanish at $u=v=0$.  Setting
$u=sx$, $v=s(1-x)$, with $0<x<1$, gives
$B_b^\eta=O(s)$ and the local integration measure
\[
 du\,dv\,W_\Delta B_b^\eta
 \ \propto\ ds\,dx\,s^{2-3\Delta}[x(1-x)]^{-\Delta}
\]
up to a bounded factor.  An unsubtracted term would instead have radial
power $s^{1-3\Delta}$ and require $\Delta<2/3$.  The subtraction removes
the common energy-independent contact term, so both radial and angular
integrals of the physical current converge for
\begin{equation}
 0<\Delta<1.
 \label{eq:B_physical_domain}
\end{equation}
Within this scaling domain, rescaling both voltage drops and temperature gives
\begin{equation}
 \mathcal I_b^\eta(\lambda\boldsymbol\varepsilon;\lambda T)
 =\lambda^{3\Delta-2}\mathcal I_b^\eta(\boldsymbol\varepsilon;T),
 \qquad \lambda>0.
 \label{eq:B_homogeneity}
\end{equation}
This simultaneous homogeneity does not by itself establish a fixed-temperature
high-bias asymptotic; the latter also requires a finite low-temperature
coefficient.

The scalar functions $F_\Delta^{\mathrm d,\mathrm u}$ introduced in
Sec.~\ref{subsec:K_directional_separation} retain the same subtraction
inside their averaged voltage brackets.  Complex conjugation followed by
$u\leftrightarrow v$, using the real symmetric $W_\Delta$ and the phase of
$A_\Delta$, gives
\begin{equation}
 F_\Delta^{\mathrm u}(E,\varepsilon_D;T)
 =e^{\mathrm{i}\pi\Delta}
  [F_\Delta^{\mathrm d}(E,\varepsilon_D;T)]^*.
 \label{eq:B_projected_conjugation}
\end{equation}
This identity supplies the equal scalar magnitudes used in
Sec.~\ref{subsec:K_inference}.  To establish that these magnitudes are
nonzero in the measurement domain, it is useful to evaluate a component
of the same subtracted integral in the energy representation.

With the Fourier convention
\begin{equation}
 g_\Delta^>(t)=\int_{-\infty}^{\infty}\frac{dE}{2\pi\hbar}
 e^{-\mathrm{i}Et/\hbar}p_\Delta(E,T),
 \label{eq:B_spectral_definition}
\end{equation}
the transform of Eq.~\eqref{eq:B_greater} is
\begin{align}
 p_\Delta(E,T)
 ={}&\frac{\tau_0^\Delta}{\Gamma(\Delta)}
  \left(\frac{2\pi T}{\hbar}\right)^{\Delta-1}e^{E/(2T)}
 \nonumber\\[-1mm]
 &\times\left|\Gamma\!\left(\frac\Delta2+\frac{\mathrm{i}E}{2\pi T}\right)\right|^2.
 \label{eq:B_spectral_density}
\end{align}
It is positive at $T>0$, obeys equilibrium detailed balance, and has the
zero-temperature limit
\begin{align}
 p_\Delta(-E,T)&=e^{-E/T}p_\Delta(E,T),
 \label{eq:B_detailed_balance}\\
 p_\Delta(E,0)&=\frac{2\pi\tau_0^\Delta}{\Gamma(\Delta)}
 \left(\frac E\hbar\right)^{\Delta-1}\Theta_{\rm H}(E).
 \label{eq:B_zero_temperature_spectrum}
\end{align}
These are scaling expressions with the infinitesimal prescription of
Eq.~\eqref{eq:B_greater}, not finite-cutoff formulas at arbitrary energies.

Choose the common drain mean as the energy origin and write
$\varepsilon_\star=|\varepsilon_D|/2$, with $E>\varepsilon_\star$.
Fourier transforming the subtracted integral, averaging the two drain
settings, and using Eq.~\eqref{eq:B_detailed_balance} gives the real part
\begin{align}
 \operatorname{Re}F_\Delta^{\mathrm d}
 ={}&-\sin(\pi\Delta)c_B\int\frac{d\omega}{2\pi\hbar}
             p_\Delta(E-\omega,T)
 \nonumber\\[-1mm]
 &\hspace{14mm}\times p_\Delta(\varepsilon_\star-\omega,T)
                         p_\Delta(\omega+\varepsilon_\star,T),
 \label{eq:B_scalar_no_zero}
\end{align}
where
\begin{align}
 c_B={}&\tfrac12(1-e^{-(E+\varepsilon_\star)/T})
 \nonumber\\[-1mm]
 &+\tfrac12e^{-2\varepsilon_\star/T}
                    (1-e^{-(E-\varepsilon_\star)/T}).
 \label{eq:B_scalar_cB}
\end{align}
The common large-negative-energy tail of the separate QPC contributions
cancels before integration, just as the common short-time term cancels in
the terminal bracket.  The resulting spectral product is exponentially
convergent at both ends.  For $T>0$ and $E>\varepsilon_\star$, $c_B$ and
all three spectral factors are strictly positive.  Thus
$\operatorname{Re}F_\Delta^{\mathrm d}<0$ for $0<\Delta<1$, in the scalar
phase convention of Sec.~\ref{subsec:K_directional_separation}; in
particular, $F_\Delta^{\mathrm d}\ne0$.  Equation~\eqref{eq:B_projected_conjugation}
then excludes a zero of $F_\Delta^{\mathrm u}$ as well.  The sign concerns
a component of a complex interference coefficient, not a positive
charge-transfer rate.

At fixed $0<\varepsilon_\star<E$, the zero-temperature spectrum restricts
the full complex scalar expression to the two intervals
$(-\varepsilon_\star,\varepsilon_\star)$ and $(\varepsilon_\star,E)$.
Their separated endpoint singularities have power $\Delta-1>-1$, so both
contributions are finite for $0<\Delta<1$.  The real part in
Eq.~\eqref{eq:B_scalar_no_zero} remains strictly negative on the first
interval.  The integrable thresholds and suppressed thermal tails also give
a finite zero-temperature limit.  Rescaling $\omega=Ew$ therefore yields
\begin{align}
 F_\Delta^{\mathrm d}(E,\rho E;T)
 ={}&\tau_0^{3\Delta}\hbar^{2-3\Delta}E^{3\Delta-2}
       f_\Delta^{\mathrm d}(\rho,T/E),
 \nonumber\\[-1mm]
 &0<|f_\Delta^{\mathrm d}(\rho,0)|<\infty,
 \qquad 0<|\rho|<2.
 \label{eq:B_fixed_detuning_scaling}
\end{align}
Here $\rho=\varepsilon_D/E$ is held away from $0$ and $\pm2$.  The
right-hand side has units of time squared, as required for the reduced
coefficient.  Equal drains are included in the finite-temperature
nonvanishing result, but not in this fixed-detuning zero-temperature limit.

\subsection{Regulated free-electron limit}
\label{app:B_contact_limit}

The point $\Delta=1$ is reached only after forming the terminal subtraction.
For a smooth bracket $B(u,v)$ with $B(0,0)=0$ and the physical large-time
regularization, the contact limit is
\begin{align}
 &\lim_{\Delta\to1^-}\sin^2(\pi\Delta)\tau_0^{3\Delta}
 \int_0^\infty du\int_0^\infty dv\,
 [uv(u+v)]^{-\Delta}B(u,v)
 \nonumber\\[-1mm]
 &\hspace{22mm}=\frac{\pi^2\tau_0^3}{3}
                  (\partial_u+\partial_v)B(0,0).
 \label{eq:B_contact_identity}
\end{align}
Only the linear term of the subtracted bracket contributes: the radial and
angular singularities cancel the vanishing $\sin^2(\pi\Delta)$ prefactor.
For the downstream bracket in Eq.~\eqref{eq:III_terminal_d},
$(\partial_u+\partial_v)B_b^{\mathrm d}(0,0)
=-3\mathrm{i}\varepsilon_b/\hbar$.
At the downstream free-electron point, the reference ordering is used
directly, with $\alpha_{\mathrm d}=1$ and $\alpha_{\mathrm u}=0$, rather
than taking a $0/0$ ratio.  Equations~\eqref{eq:B_factorization} and
\eqref{eq:B_contact_identity} give the collected-current coefficient
\begin{equation}
 \lim_{\Delta\to1^-}\mathcal I_b
 =\frac{4\mathrm{i}\pi^2\tau_0^3}{\hbar}\varepsilon_b.
 \label{eq:B_contact_current}
\end{equation}

To fix the scattering normalization, use canonical chiral fermions with
short-distance lengths $a_c=v_c\tau_0$.  A local hopping amplitude
$\gamma_{ba}$ in $\gamma_{ba}\psi_b^\dagger\psi_a+\mathrm{H.c.}$ corresponds
to $\Gamma_{ba}=\gamma_{ba}/(2\pi\sqrt{a_ba_a})$.  Its Born amplitude
between flux-normalized channels is therefore
\begin{equation}
 t_{ba}=-\frac{\mathrm{i}\gamma_{ba}}{\hbar\sqrt{v_bv_a}}
       =-\mathrm{i}\frac{2\pi\tau_0}{\hbar}\Gamma_{ba}.
 \label{eq:B_t_Gamma}
\end{equation}
The product of the three scattering amplitudes, including the flux and
accumulated free propagation phase $\varphi_{\rm prop}(E)$, is
\begin{equation}
 \Lambda(E)=\mathrm{i}\left(\frac{2\pi\tau_0}{\hbar}\right)^3
 e^{\mathrm{i}[2\pi e\Phi/h+\varphi_{\rm prop}(E)]}
 \Gamma_{02}\Gamma_{21}\Gamma_{10}.
 \label{eq:B_Lambda}
\end{equation}
With the harmonic normalization \eqref{eq:II_harmonic_definition}, the
contact coefficient reproduces the Landauer form used in
Eq.~\eqref{eq:K_equal_drain_fermion}:
\begin{equation}
 I_b^{(3),\mathrm{AB}}
 =\frac{2e}{h}\int dE\,\operatorname{Re}\Lambda(E)
                [f_b(E)-f_{b+1}(E)].
 \label{eq:B_Landauer}
\end{equation}
Here $f_c$ is the Fermi occupation of states incident from reservoir $c$.
For common temperature and an energy-independent loop amplitude in the
unresolved-flight limit, the occupation integral equals
$\varepsilon_b=e(V_b-V_{b+1})$, giving Eq.~\eqref{eq:B_contact_current}
with its absolute normalization.  Equal drain occupations then cancel this
harmonic as described in Sec.~\ref{subsec:K_equal_drains}.

\section{Gaussian derivation of the interaction response}
\label{app:bridge}
\label{app:single_bridge}

We derive the dark-terminal response of Sec.~\ref{sec:bridge} and the leading
bright scaling in the same device.  The density interaction is included in
the equilibrium Gaussian correlations before expanding in QPC tunneling.
This gives a common starting point for both currents.  We then expand to
first order for the spatial coefficient and the bright infrared limit.
Throughout, only $C_a$ is interacting, and the other two sectors remain local
downstream Laughlin edges with unresolved flight times.

\subsection{Gaussian correlations and terminal response}
\label{app:C_preparation}
\label{app:C_pairs}

Use the interaction $H_{U,a}=\lambda_a\mathcal U_a$ of
Eq.~\eqref{eq:bridge_HU}, the source $a+1$ and equal drains $a-1,a$ of
Eq.~\eqref{eq:bridge_bias}, and the collected-current convention of
Appendix~\ref{app:cubic_current}.  The independent sectors are at a common
temperature, and the quadratic Hamiltonian is stable.  We abbreviate the
selected Laughlin vertex as $\mathcal V_c=:\!e^{\mathrm{i}\Phi_c}\!:$,
with $q=\nu e$.  In the voltage-phase representation the equilibrium reference is
\begin{equation}
 \mathcal K_{\lambda_a}=K_0+\lambda_a\mathcal U_a,
 \qquad
 \rho_{\lambda_a}=Z_{\lambda_a}^{-1}e^{-\mathcal K_{\lambda_a}/T}.
 \label{eq:C_Gibbs_reference}
\end{equation}
Since $[Q_b,H_{U,a}]=0$, the current operator is unchanged.  The cubic
expansion \eqref{eq:A_nested_current} is evaluated with the evolution and
expectation value generated by this reference.

For an ordinary vertex pair at space-time points $1,2$, Gaussian averaging
gives
\begin{equation}
 \frac{G_{a,\lambda_a}(1,2)}{G_{a,0}(1,2)}
 =\exp\!\left[
 \Delta_{\lambda_a}\mathsf C_a(1,2)
 -\frac12\sum_{j=1}^2\Delta_{\lambda_a}\mathsf C_a(j,j)
 \right].
 \label{eq:C_gaussian_pair}
\end{equation}
Here $\mathsf C_a$ is the covariance of the vertex phase $\Phi_a$ in the
specified operator order, and
$\Delta_{\lambda_a}\mathsf C_a=\mathsf C_{a,\lambda_a}-\mathsf C_{a,0}$ is
an exact difference, not a first-order variation.  A common regulator and
fixed vertex normalization are understood.  The equal-point terms include
the change in equilibrium fluctuations; they are retained when this formula
is expanded.  No separate expansion of the equilibrium state is needed.

Define the endpoint pairs
\begin{align}
 G_c^>(t)&=\langle\mathcal V_c(x_{c,d},t)
                     \mathcal V_c^\dagger(x_{c,u},0)\rangle_{\lambda_a},
 \nonumber\\
 G_c^<(t)&=\langle\mathcal V_c^\dagger(x_{c,u},0)
                     \mathcal V_c(x_{c,d},t)\rangle_{\lambda_a},
 \label{eq:C_pairs}
\end{align}
and the combination
\begin{equation}
 \zeta=e^{\mathrm{i}\pi\nu},\qquad
 \mathcal B_c(t)=G_c^>(t)-\zeta G_c^<(t).
 \label{eq:C_B_definition}
\end{equation}
Only the pairs on $C_a$ depend on $\lambda_a$.  On the two local spectator
edges, $\mathcal B_{a\pm1}(t)=0$ for $t<0$; finite positive flight-time
thresholds are fixed before their delays are neglected.  With
$G(\omega)=\int dt\,e^{\mathrm{i}\omega t/\hbar}G(t)$, where $\omega$ is an
energy, equilibrium detailed balance gives
\begin{equation}
 G_c^<(\omega)=e^{-\omega/T}G_c^>(\omega).
 \label{eq:C_balance}
\end{equation}

For $u=-t_1>0$ and $v=t_1-t_2>0$, expanding the nested commutators
\eqref{eq:A_Md}--\eqref{eq:A_Mu} and regrouping vertices by edge gives
\begin{align}
 M_c^{\mathrm d}(u,v)
 ={}&\mathcal B_{c-1}(u)
 \bigl[\mathcal B_{c+1}(v)G_c^<(-u-v)
 \nonumber\\[-1mm]
 &\hspace{17mm}-G_{c+1}^<(v)\mathcal B_c(-u-v)\bigr],
 \label{eq:C_pair_Md}\\
 M_c^{\mathrm u}(u,v)
 ={}&-\mathcal B_c(-u)
 \bigl[G_{c+1}^<(-v)\mathcal B_{c-1}(u+v)
 \nonumber\\[-1mm]
 &\hspace{17mm}-\mathcal B_{c+1}(-v)G_{c-1}^<(u+v)\bigr].
 \label{eq:C_pair_Mu}
\end{align}
The relative exchange phase derived in
Eq.~\eqref{eq:B_common_ordered_product} is already included.  Each term
contains one pair from each edge.  The full terminal current is therefore a
linear functional of the interacting pair, although that pair need not be
linear in $\lambda_a$.

Insert these expressions into the voltage-weighted integrals
\eqref{eq:A_Jd}--\eqref{eq:A_Ju} and form the terminal differences
\eqref{eq:A_terminal_components}.  Taking the common drain potential as the
energy origin, the dark-drain contribution obtained by extending the first term of $M_c^{\mathrm d}$ to the full
$(u,v)$ plane contains the spectral bracket
\begin{equation}
 G_{a-1}^<(\omega)\mathcal B_a(\omega)
 -G_a^<(\omega)\mathcal B_{a-1}(\omega)=0.
 \label{eq:C_equal_drain_balance}
\end{equation}
The equality follows from
$\mathcal B_c(\omega)=(e^{\omega/T}-\zeta)G_c^<(\omega)$ at the two equal
voltage drains.  It does not require identical drain spectra.  The added
negative-time domains, however, must still be combined with the other terms
of Eqs.~\eqref{eq:C_pair_Md}--\eqref{eq:C_pair_Mu}.

At fixed $r>0$, the completion term has $t>0$ and contains
$\mathcal B_{a+1}(t)G_{a-1}^<(r-t)$.  The two time orderings supply the
opposite-signed product $G_{a+1}^<(t)\mathcal B_{a-1}(r-t)$ on
$0<t<r$ and $t<0$.  The spectator support permits both products to be
extended to the real line.  Their difference is
\begin{align}
 \mathcal C(r,E)
 ={}&\int_{-\infty}^{\infty}dt\,e^{-\mathrm{i}Et/\hbar}
 \bigl[G_{a+1}^>(t)G_{a-1}^<(r-t)
 \nonumber\\[-1mm]
 &\hspace{25mm}-G_{a+1}^<(t)G_{a-1}^>(r-t)\bigr]
 \nonumber\\
 ={}&-(1-e^{-E/T})\int\frac{d\omega}{2\pi\hbar}\,
 e^{-\mathrm{i}\omega r/\hbar}
 \nonumber\\[-1mm]
 &\hspace{25mm}\times p_\nu(\omega,T)p_\nu(E-\omega,T).
 \label{eq:C_exterior_time}
\end{align}
Here $p_\nu$ is the local spectrum \eqref{eq:B_spectral_density} at exponent
$\nu$.  The terms proportional to $\zeta G^<G^<$ cancel in the first line;
the last line follows by Fourier transformation and detailed balance.
Thus the complete dark coefficient is
\begin{equation}
 \mathcal I_{a-1}[\lambda_a]
 =-\mathrm{i}\int_0^\infty dr\,
       \mathcal B_a(-r)\mathcal C(r,E).
 \label{eq:C_gaussian_dark_current}
\end{equation}

For the bright drain, the change of variable $t\mapsto r-t$ in the source
contribution combines the ranges $t<0$ and $0<t<r$ into the same integral
as at the neighboring current link.  These additional domains cancel in
$\mathcal J_{a+1}-\mathcal J_a$.  The remaining spectral difference gives
\begin{align}
 \mathcal I_a[\lambda_a]
 ={}&-\mathrm{i}(1-e^{-E/T})\int\frac{d\omega}{2\pi\hbar}\,
 G_a^>(\omega)
 \nonumber\\[-1mm]
 &\times\bigl[p_\nu(\omega,T)-\zeta p_\nu(-\omega,T)\bigr]
 p_\nu(E-\omega,T).
 \label{eq:C_gaussian_bright_current}
\end{align}
The source coefficient is $-\mathcal I_{a-1}-\mathcal I_a$.  Both expressions
vanish at full equilibrium.  At zero interaction, the negative-time
$\mathcal B_a$ is zero, and Eq.~\eqref{eq:C_gaussian_bright_current} reduces
to $F_\nu^{\mathrm d}(E,0;T)$ after the same unresolved-flight limit.
Equations~\eqref{eq:C_gaussian_dark_current} and
\eqref{eq:C_gaussian_bright_current} keep the full Gaussian pair on one
edge at cubic QPC order.  They include both time orderings and the
normalized equilibrium correlations, rather than only a retarded response.

\subsection{Weak spatial interaction and the static dark signal}
\label{app:C_support}
\label{app:C_static}

The negative-time pair difference has a simple Gaussian interpretation.
Let $\kappa_a^{\mathrm u}(r)=[\Theta_a(0),\Theta_{a+1}(-r)]$ be the
c-number commutator of the QPC phases in Eq.~\eqref{eq:III_Theta}, with
the same fixed cross-edge order as in Appendix~\ref{app:vertex_loop}.
It vanishes in the local downstream reference.  The exchange relation gives
\begin{equation}
 \mathcal B_a(-r)
 =\zeta G_a^<(-r)\bigl[e^{\kappa_a^{\mathrm u}(r)}-1\bigr].
 \label{eq:C_gaussian_reverse_pair}
\end{equation}
For $\delta X=\lambda_a(\partial_{\lambda_a}X)_0$, its first variation is
\begin{align}
 K_a(r)&\equiv\delta\mathcal B_a(-r)
       =\zeta G_{a,0}^<(-r)\delta\kappa_a^{\mathrm u}(r),
 \nonumber\\[-1mm]
 G_{a,0}^<(-r)&=g_\nu^>(r+D_a/v_a).
 \label{eq:C_K_kappa}
\end{align}
The variation of the common magnitude multiplies the vanishing bare reverse
response.  It remains included in the ordinary pair in
Eq.~\eqref{eq:C_gaussian_bright_current}.

The algebra $[\Phi_a(x),\Phi_a(y)]=\mathrm{i}\pi\nu\operatorname{sgn}(x-y)$
and the interaction \eqref{eq:bridge_HU} give
\begin{equation}
 (\partial_t+v_a\partial_x)\Phi_a(x,t)
 =-\frac{\lambda_a\nu}{2\pi\hbar}
       \int dz\,U_a(x,z)\partial_z\Phi_a(z,t).
 \label{eq:C_field_equation}
\end{equation}
Integrating along an unperturbed downstream characteristic yields
\begin{align}
 \delta\kappa_a^{\mathrm u}(r)
 ={}&\frac{\mathrm{i}\lambda_a\nu^2}{\hbar}\int_0^r d\tau\,
 \nonumber\\[-1mm]
 &\times U_a\bigl(x_{a,u}-v_a(r-\tau),
                  x_{a,d}+v_a\tau\bigr).
 \label{eq:C_kappa}
\end{align}
Both propagation legs run downstream.  The interaction can join them only
across $y<x_{a,u}<x_{a,d}<z$, the support illustrated in
Fig.~\ref{fig:static_bridge}.  A local kernel, or one with no such support,
therefore produces no dark signal at first interaction order.

The static coefficient is obtained from $C_{a,0}(T)=\int_0^\infty dr\,K_a(r)$.
Changing variables gives
\begin{align}
 y&=x_{a,u}-v_a(r-\tau),\qquad z=x_{a,d}+v_a\tau,
 \nonumber\\
 dr\,d\tau&=\frac{dy\,dz}{v_a^2},\qquad
 r+\frac{D_a}{v_a}=\frac{z-y}{v_a}.
 \label{eq:C_spatial_map}
\end{align}
Using the phase and thermal envelope of
Eq.~\eqref{eq:B_thermal_envelope}, the moment is
\begin{align}
 C_{a,0}(T)
 ={}&\frac{\mathrm{i}\lambda_a\nu^2}{\hbar v_a^2}
 e^{\mathrm{i}\pi\nu/2}
 \int_{-\infty}^{x_{a,u}}dy\int_{x_{a,d}}^\infty dz
 \nonumber\\[-1mm]
 &\hspace{12mm}\times U_a(y,z)
 \mathcal E_\nu\!\left(\frac{z-y}{v_a},T\right).
 \label{eq:C_thermal_moment}
\end{align}
The full separation $z-y$ enters the vertex factor, whereas the extra
response time is $(z-y-D_a)/v_a$.

If the external energies do not resolve this extra time,
$\mathcal C(r,E)\simeq\mathcal C(0,E)=-P_\nu^{\rm net}(E,T)$, with
$P_\nu^{\rm net}$ defined in Eq.~\eqref{eq:bridge_Pnet}.  The linear dark
coefficient becomes
\begin{equation}
 \delta\mathcal I_{a-1}^{\rm stat}
 =\mathrm{i}C_{a,0}(T)P_\nu^{\rm net}(E,T).
 \label{eq:C_static_thermal}
\end{equation}
When also $T(z-y)/(\hbar v_a)\ll1$ over appreciable support, the envelope in
Eq.~\eqref{eq:C_thermal_moment} reduces to $(v_a\tau_0/(z-y))^\nu$.
The spatial integral is then $\mathcal Q_{\nu,a}[U_a]$ of
Eq.~\eqref{eq:bridge_Qmoment}, giving
\begin{equation}
 \begin{aligned}
 \delta\mathcal I_{a-1}&\simeq
 -\frac{\lambda_a\nu^2}{\hbar v_a^2}\mathcal Q_{\nu,a}[U_a]\\
 &\quad\times e^{\mathrm{i}\pi\nu/2}P_\nu^{\rm net}(E,T).
 \end{aligned}
 \label{eq:C_static_current}
\end{equation}
This is Eq.~\eqref{eq:bridge_dark_master}, with the sign fixed by positive
collected current.  The two exterior spectra determine its leading
$E^{2\nu-1}$ amplitude for $E\gg T$.

The two small-time requirements are distinct and are stated in
Eq.~\eqref{eq:bridge_unresolved_condition}.  The contributing distances must
also be within the physical edge-scaling regime; $\tau_0$ fixes operator
normalization and is not a prescribed microscopic cutoff.  A zero signed
moment removes only the leading static term, so higher response moments may
matter near a cancellation.  These limits are taken within the weak-QPC
window \eqref{eq:IV_bridge_window}, with temperature above the crossover of
the unbiased contact.

\subsection{Bright scaling in the interacting device}
\label{app:C_reference}

Unlike the dark coefficient, Eq.~\eqref{eq:C_gaussian_bright_current}
contains the ordinary pair $G_a^>$ and hence its interaction-dependent
normalization.  For a weak kernel localized near the QPC region, with
finite spatial moments and a regular single-mode infrared limit, its
long-time exponent remains $\nu$.  To see the origin of this statement,
consider a mode $f_\omega(x)e^{-\mathrm{i}\omega t/\hbar}$ with unit incoming
amplitude.  With $k=\omega/(\hbar v_a)$, the first-order solution of
Eq.~\eqref{eq:C_field_equation} is
\begin{align}
 f_\omega(x)
 ={}&e^{\mathrm{i}kx}\left[1-
 \frac{\mathrm{i}\lambda_a\nu k}{2\pi\hbar v_a}
 \int_{-\infty}^{x}dy\int dz\,U_a(y,z)
 e^{\mathrm{i}k(z-y)}\right]
 \nonumber\\[-1mm]
 &+O(\lambda_a^2).
 \label{eq:C_infrared_mode}
\end{align}
For real $U_a$ the leading correction is imaginary and linear in $k$.
The zero-frequency mode weight is unchanged, so the coefficient of the
infrared logarithm in the Gaussian covariance is still $\nu$.  The finite
covariance change instead supplies a constant factor $Z_a=1+O(\lambda_a)$:
\begin{equation}
 G_a^>(t;\lambda_a)
 =Z_a\,g_\nu^>(t)+\text{subleading low-energy terms}.
 \label{eq:C_bright_infrared_pair}
\end{equation}
Here the leading normalization is independent of bias and temperature when
the relevant scales of the localized interaction are unresolved.  This
condition concerns the full profile, not only the cross-region moment
entering the dark signal.

The local function obtained by setting the interaction to zero in
Eq.~\eqref{eq:C_gaussian_bright_current} can equivalently be written
\begin{align}
 F_\nu^{\mathrm d}(E,0;T)
 ={}&A_\nu\int_0^\infty du\int_0^\infty dv\,W_\nu(u,v;T)
 \nonumber\\[-1mm]
 &\times\bigl[e^{\mathrm{i}E(u+v)/\hbar}
                 -e^{-\mathrm{i}Ev/\hbar}\bigr].
 \label{eq:C_bright_integral}
\end{align}
This is the same scalar function as in Sec.~\ref{subsec:K_directional_separation},
with the subtraction retained.  For $0<\nu<1/2$ and $T\ll E$, the dominant
region has $v\sim\hbar/E$ and $u\sim\hbar/T$, with envelope
$(\tau_0/v)^\nu\mathcal E_\nu(u,T)^2$.  The standard integrals are
\begin{align}
 \int_0^\infty dv\,v^{-\nu}e^{-\mathrm{i}Ev/\hbar}
 &=\Gamma(1-\nu)e^{-\mathrm{i}\pi(1-\nu)/2}(\hbar/E)^{1-\nu},
 \nonumber\\
 \int_0^\infty du\,\mathcal E_\nu(u,T)^2
 &=\tau_0^{2\nu}\left(\frac{\pi T}{\hbar}\right)^{2\nu-1}
 \nonumber\\[-1mm]
 &\quad\times\frac{2^{2\nu-1}\Gamma(\nu)\Gamma(1-2\nu)}{\Gamma(1-\nu)}.
 \label{eq:C_bright_endpoints}
\end{align}
With $A_\nu$ from Eq.~\eqref{eq:B_factorization}, this gives
\begin{align}
 F_\nu^{\mathrm d}(E,0;T)
 ={}&-2^{2\nu+1}\sin^2(\pi\nu)\Gamma(\nu)\Gamma(1-2\nu)
 \nonumber\\[-1mm]
 &\times\tau_0^{3\nu}\hbar^{2-3\nu}
 E^{\nu-1}(\pi T)^{2\nu-1}e^{\mathrm{i}\pi\nu}
 \nonumber\\[-1mm]
 &\times\left[1+O\!\left((T/E)^{1-2\nu}\right)\right].
 \label{eq:C_bright_asymptotic}
\end{align}
The regular two-time region supplies the stated relative correction.  The
minus sign follows from the current scalar subtraction in
Eq.~\eqref{eq:C_bright_integral}.

Finally, linearity of Eq.~\eqref{eq:C_gaussian_bright_current} in $G_a^>$
and Eq.~\eqref{eq:C_bright_infrared_pair} give, at the retained weak-interaction
and infrared order,
\begin{align}
 \mathcal I_a[\lambda_a]
 ={}&Z_aF_\nu^{\mathrm d}(E,0;T)
 \nonumber\\[-1mm]
 &+\text{subleading low-energy terms}+O(\lambda_a^2).
 \label{eq:C_bright_renormalized}
\end{align}
Thus the measured bright amplitude retains the leading powers
$E^{\nu-1}T^{2\nu-1}$, although its prefactor need not be the local one.
It can serve as a scaling reference without switching off a natural
interaction.  This conclusion is restricted to the regular weak localized
single-mode regime above, with the QPC magnitudes fixed or known.  It neither
identifies the full interacting amplitude with the bare reference nor makes
the local directional ratio of Sec.~\ref{sec:kmatrix} an intrinsic
reconstruction for an arbitrary interacting device.

\end{document}